\documentstyle[preprint,aps,epsf,floats]{revtex}

\tighten

\def\OMIT#1{{}}

\def\Dslash{D\hskip-0.65em /}
\def\Se{s}
\def\Tr{t}
\def\SD{x}
\def\SS{s}
\def\ST{t}
\def\SD{x}
\def\VS{\tilde s}
\def\VT{\tilde t}

\def\cb{{\cal B}}
\def\cbb{{\overline{\cal B}}}
\def\cf{{\cal F}}

\begin{document}

\preprint{\vbox{
\hbox{NT@UW-02-004}
}}

\title{Nucleons
in Two-Flavor Partially-Quenched\\ Chiral Perturbation Theory}
\author{{\bf Silas R. Beane}  and {\bf Martin J. Savage}}
\address{Department of Physics, University of Washington, \\
Seattle, WA 98195. }

\maketitle

\begin{abstract} 
  
  Properties of the proton and the neutron are explored in partially-quenched
  chiral perturbation theory with two non-degenerate light flavors.  Masses,
  magnetic moments, matrix elements of isovector twist-2 operators and
  axial-vector currents are computed at the one-loop level in the chiral
  expansion.

\end{abstract}

\vfill\eject

\section{Introduction}

At some point in time lattice calculations will be able to determine the
properties and interactions of the low-lying hadrons directly from QCD.
Impressive progress is being made toward this goal, however present
computational limitations necessitate the use of quark masses that are
significantly larger than those of nature.  Typically one has 
$m_\pi^{\rm latt.}\sim 500~{\rm MeV}$.  In order to make a connection between lattice
calculations of the foreseeable future and nature, an extrapolation in the
quark masses is required.  Chiral perturbation theory ($\chi$PT) provides a
systematic description of low-energy QCD near the chiral limit, and this
technique has been extended to describe both quenched QCD
(QQCD)~\cite{Sharpe90,S92,BG92,LS96,S01a} and partially-quenched QCD
(PQQCD)~\cite{Pqqcd,SS01}.  The hope is that future lattice simulations can be
performed with quark masses that are small enough to guarantee a convergent
chiral expansion, thus allowing a meaningful extrapolation to the quark masses of
nature.

Recently, it has been shown how to include the low-lying octet and
decuplet of baryons~\cite{CS01a} into partially-quenched chiral perturbation theory
(PQ$\chi$PT)~\cite{Pqqcd,SS01}.  As the pion mass presently obtainable in
lattice simulations is not much smaller than the physical kaon mass, it is
natural to consider a theory with three light valence-quarks and three light
sea-quarks.  The masses, magnetic moments and matrix elements of isovector
twist-2 operators were computed at the one-loop level in 
$SU(6|3)_L\otimes SU(6|3)_R$ PQ$\chi$PT~\cite{CS01a}. 
The strange quark mass was taken to be different from
the other two degenerate light quarks (isospin limit) and one of the sea-quarks
was taken to have a mass different from the other two.  However, many lattice
simulations are performed with just two light quarks.  In this work we
determine several observables of the proton and the neutron at one-loop level in
PQ$\chi$PT with two non-degenerate light quarks using 
$SU(4|2)_L\otimes SU(4|2)_R$ PQ$\chi$PT. Isospin violation of electromagnetic origin is not
considered.

In Section~\ref{sec:formalism} we develop the general formalism for including
baryons in $SU(4|2)_L\otimes SU(4|2)_R$ PQ$\chi$PT.  The masses and magnetic
moments of the proton and neutron are computed to one-loop order in the chiral
expansion in Section~\ref{sec:masses} and~\ref{sec:mm}, respectively. In
Section~\ref{sec:for} and~\ref{sec:axial} we present one-loop level expressions
for matrix elements of isovector twist-2 operators and axial-vector currents,
respectively.  We conclude in Section~\ref{sec:conc}.

\section{PQ$\chi$PT}
\label{sec:formalism}

The Lagrange density describing the quark-sector of PQQCD is
\begin{eqnarray}
{\cal L} & = & 
\sum_{a,b=u,d} \overline{q}_V^a
\left[\ i\Dslash -m_{q}\ \right]_a^b\ q_{V b}
\ +\ 
\sum_{\tilde a,\tilde b=\tilde u, \tilde d}
\overline{\tilde q}^{\tilde a}
\left[\ i\Dslash -m_{\tilde q}\ 
\right]_{\tilde a}^{\tilde b}\tilde q_{\tilde b}
\ +\ 
\sum_{a,b=j,l} \overline{q}_{\rm sea}^a
\left[\ i\Dslash -m_{\rm sea}\ \right]_a^b\ q_{{\rm sea}, b}
\nonumber\\
& = & 
\sum_{k,n=u,d,\tilde u, \tilde d,j,l} 
\overline{Q}^k\ 
\left[\ i\Dslash -m_{Q}\ \right]_k^n\ Q_n
\ \ \ ,
\label{eq:PQQCD}
\end{eqnarray}
where the $q_V$ are the two light-quarks, $u$ and $d$,
the $\tilde q$ are two light bosonic (ghost) quarks  
$\tilde u$ and  $\tilde d$,
and the $q_{{\rm sea}}$ are the two sea-quarks $j$ and  $l$.
The left- and right-handed 
valence-, sea-, and ghost-quarks are combined into
column vectors
\begin{eqnarray}
Q_L & = & \left(u,d,j,l,\tilde u,\tilde d\right)^T_L
\ \ ,\ \ 
Q_R \ =\  \left(u,d,j,l,\tilde u,\tilde d\right)^T_R
\ \ \ ,
\label{eq:quarkvec}
\end{eqnarray}
where the graded equal-time commutation relations for two fields is
\begin{eqnarray}
Q_i^\alpha ({\bf x}) Q_k^{\beta \dagger} ({\bf y}) - 
(-)^{\eta_i\eta_k}Q_k^{\beta \dagger} ({\bf y})Q_i^\alpha ({\bf x})
& = & 
\delta^{\alpha\beta}\delta_{ik}\delta^3({\bf x}-{\bf y})
\ \ \ ,
\label{eq:comm}
\end{eqnarray}
where $\alpha,\beta$ are spin-indices and $i,k$ are flavor indices.
The objects $\eta_k$ correspond to the parity of the component of $Q_k$,
with $\eta_k=+1$ for $k=1,2,3,4$ and $\eta_k=0$ for $k=5,6$, and 
the graded commutation relations for two $Q$'s or two $Q^\dagger$'s
are analogous. 
The $Q_{L,R}$ in eq.~(\ref{eq:quarkvec})
transform in the fundamental representation of $SU(4|2)_{L,R}$
respectively. 
The ground floor of $Q_L$ transforms as a $({\bf 4},{\bf 1})$ of
$SU(4)_{q L}\otimes SU(2)_{\tilde q L}$ while the first floor transforms
as $({\bf 1},{\bf 2})$, and the 
right handed field $Q_R$ transforms analogously.

In the absence of quark masses, $m_Q=0$,
the Lagrange density in eq.~(\ref{eq:PQQCD})
has a
graded symmetry $U(4|2)_L\otimes U(4|2)_R$, where the left- and 
right-handed quark fields transform as
$Q_L\rightarrow U_L Q_L$ and $Q_R\rightarrow U_R Q_R$ respectively.
The strong anomaly reduces the symmetry of the theory, which can be
taken to be 
$SU(4|2)_L\otimes SU(4|2)_R\otimes U(1)_V$~\cite{SS01}.
It is assumed that this symmetry is spontaneously broken according to the pattern
$SU(4|2)_L\otimes SU(4|2)_R\otimes U(1)_V\rightarrow SU(4|2)_V\otimes U(1)_V$ 
so that an identification with QCD can be made.

The mass-matrix, $m_Q$, has entries
$m_Q = {\rm diag}(m_u,m_d,m_j,m_l,m_u,m_d)$,
i.e.  
$m_{\tilde u}=m_u$ and  $m_{\tilde d}=m_d$, so that the contribution 
to the determinant in the path integral from integrating over 
the $q_V$ and the $\tilde q$ exactly cancel, leaving the 
contribution from the $q_{\rm sea}$ alone.

\subsection{The Pseudo-Goldstone Bosons}

The strong interaction dynamics of the 
pseudo-Goldstone bosons are described  at leading order (LO)
in PQ$\chi$PT by a Lagrange density of the form~\cite{Pqqcd},
\begin{eqnarray}
{\cal L } & = & 
{f^2\over 8} 
{\rm str}\left[\ \partial^\mu\Sigma^\dagger\partial_\mu\Sigma\ \right]
 \ +\ 
\lambda\ {f^2\over 4} 
{\rm str}\left[\ m_Q\Sigma^\dagger + m_Q\Sigma\ \right]
\ +\ 
\alpha_\Phi\partial^\mu\Phi_0\partial_\mu\Phi_0\ -\ m_0^2\Phi_0^2
\ \ \ \ ,
\label{eq:lagpi}
\end{eqnarray}
where $\alpha_\Phi$ and $m_0$ are quantities that do not vanish in the 
chiral limit.
The meson field is incorporated in $\Sigma$ via
\begin{eqnarray}
\Sigma & = & \exp\left({2\ i\ \Phi\over f}\right)
\ =\ \xi^2
\ \ \ ,\ \ \ 
\Phi \ =\  \left(\matrix{ M &\chi^\dagger \cr \chi &\tilde{M} }\right)
\ \ \ ,
\label{eq:phidef}
\end{eqnarray}
where $M$ and $\tilde M$ are matrices containing bosonic mesons while
$\chi$ and $\chi^\dagger$ are matrices containing fermionic mesons,
with
\begin{eqnarray}
M & = & 
\left(\matrix{
\eta_u & \pi^+ & J^0 & L^+ \cr
\pi^- & \eta_d & J^- & L^0\cr
\overline{J}^0 & J^+ & \eta_j & Y_{jl}^+\cr
L^- & \overline{L}^0 & Y_{jl}^- & \eta_l  }
\right)
\ ,\ 
\tilde M \ =\  \left(\matrix{\tilde\eta_u & \tilde\pi^+ \cr
\tilde\pi^- & \tilde\eta_d
}\right)
\ ,\
\chi \ =\ 
\left(\matrix{\chi_{\eta_u} & \chi_{\pi^+} &  
\chi_{J^0} & \chi_{L^+}\cr
\chi_{\pi^-} & \chi_{\eta_d} & 
\chi_{J^-} & \chi_{L^0} }
\right)
\ ,
\label{eq:mesdef}
\end{eqnarray}
where the upper $2\times 2$ block of $M$ is the usual triplet plus singlet 
of pseudo-scalar mesons while the remaining entries correspond to mesons
formed with the sea-quarks.
The convention we use corresponds to $f~\sim~132~{\rm MeV}$,
and the charge assignments have been made using an electromagnetic charge
matrix,  ${\cal Q}^{(PQ)} = {1\over 3} {\rm diag}\left(2,-1,2,-1,2,-1\right)$.

The singlet field is defined to be 
$\Phi_0 ={\rm str}\left(\ \Phi\ \right)/\sqrt{2}$,
and its mass $m_0$ can be taken to
be of order the scale of chiral symmetry breaking, 
$m_0\rightarrow\Lambda_\chi$~\cite{SS01}.
In taking this limit, one finds that the $\eta$
two-point functions deviate from simple single-poles.
The $\eta_a\eta_b$ propagator is found to be
\begin{eqnarray}
{\cal G}_{\eta_a\eta_b} & = & 
{ i \delta^{ab}\over q^2- m_{\eta_a}^2 + i \epsilon}
\ -\ {i\over 2}
{(q^2-m_{jj}^2)(q^2-m_{ll}^2)\over (q^2- m_{\eta_a}^2 + i \epsilon)
(q^2- m_{\eta_b}^2 + i \epsilon)
(q^2- {1\over 2}(m_{jj}^2+m_{ll}^2)+ i \epsilon)}
\ \ ,
\end{eqnarray}
where $m_{ab}$ is the mass of the meson composed of (anti)-quarks of flavor
$a$ and $b$.
This can be compactly written as
\begin{eqnarray}
{\cal G}_{\eta_a\eta_b} & = &  \delta^{ab}  P_a\ +\ 
{\cal H}_{ab}( P_a , P_b , P_X)
\ \ ,
\label{eq:HPs}
\end{eqnarray}
where 
\begin{eqnarray}
& & P_a\ =\  { i \over q^2- m_{\eta_a}^2 + i \epsilon}
\ \ \ ,\ \ \ 
P_b \ = \  { i \over q^2- m_{\eta_b}^2 + i \epsilon}
\ \ \ ,\ \ \ 
P_X \ = \  { i \over q^2- m_X^2 + i \epsilon}
\nonumber\\
& & {\cal H}_{ab}( A, B, C) \ =\  
-{1\over 2}\left[\ 
{(m_{jj}^2-m_{\eta_a}^2)(m_{ll}^2-m_{\eta_a}^2)\over 
(m_{\eta_a}^2-m_{\eta_b}^2)(m_{\eta_a}^2-m_X^2)}\  A
-
{(m_{jj}^2-m_{\eta_b}^2)(m_{ll}^2-m_{\eta_b}^2)\over 
(m_{\eta_a}^2-m_{\eta_b}^2)(m_{\eta_b}^2-m_X^2)}\  B
\right.\nonumber\\ & & \left.\qquad
\ +\ 
{(m_X^2-m_{jj}^2)(m_X^2-m_{ll}^2)\over 
(m_X^2-m_{\eta_a}^2)(m_X^2-m_{\eta_b}^2)}\  C
\ \right]
\ \ \ ,
\label{eq:HPsdef}
\end{eqnarray}
where the mass, $m_X$, is given by 
$m_X^2 = {1\over 2}\left( m_{jj}^2+ m_{ll}^2 \right)$.
One important implication of this form for the singlet propagator is that
the nucleon-nucleon potential falls off exponentially at long distances, 
as opposed to Yukawa-like behavior, away from the QCD-limit~\cite{BS02a}.

The tree-level masses resulting from the Lagrange density in 
eq.~(\ref{eq:lagpi}) are
\begin{eqnarray}
m_{\pi^\pm}^2 & = & m_{ud}^2\ =\ 
\lambda\ \left(m_u+m_d\right)
\ \ ,\ \ 
m_{\eta_u}^2 \ =\  m_{uu}^2\ =\ 
2\ \lambda\  m_u
\ \ ,
\nonumber\\
m_{J^0}^2 & = &  m_{ju}^2\ =\ 
\lambda\ \left(m_j+m_u\right)
\ \ \ ,
\end{eqnarray}
and so forth.

\subsection{The Nucleons and $\Delta$-Resonances}

It is straightforward to include the proton, neutron, 
and $\Delta$-resonances into PQ$\chi$PT~\cite{LS96,CS01a}.
An interpolating field that has non-zero overlap with the 
nucleons (when the $ijk$ indices are restricted to $1,2$) 
is~\cite{LS96}
\begin{eqnarray}
{\cal B}^\gamma_{ijk} & \sim &
\left[\ Q_i^{\alpha,a} Q_j^{\beta,b} Q_k^{\gamma,c}
\ -\  Q_i^{\alpha,a} Q_j^{\gamma,c} Q_k^{\beta,b}\ \right]
\epsilon_{abc} \left(C\gamma_5\right)_{\alpha\beta}
\ \ \ ,
\label{eq:octinter}
\end{eqnarray}
where $C$ is the charge conjugation operator,
$a,b,c$ are color indices and $\alpha,\beta,\gamma$ are Dirac indices.
Dropping the Dirac index, one finds that
under the interchange of flavor indices~\cite{LS96},
\begin{eqnarray}
{\cal B}_{ijk} & = & (-)^{1+\eta_j \eta_k}\  {\cal B}_{ikj}
\ \ ,\ \ 
{\cal B}_{ijk} \ +\  (-)^{1+\eta_i \eta_j}\ {\cal B}_{jik}
\ +\ (-)^{1 + \eta_i\eta_j + \eta_j\eta_k + \eta_k\eta_i}\ 
{\cal B}_{kji}\ =\ 0
\ \ \ .
\label{eq:bianchi}
\end{eqnarray}
In analogy with QCD, 
we consider the transformation of ${\cal B}_{ijk}$ under $SU(4|2)_V$
transformations, and using the graded relation
\begin{eqnarray}
Q_i\ U^j_{\ k} & = & (-)^{\eta_i (\eta_j+\eta_k)} \ U^j_{\ k}\  Q_i
\ \ ,
\label{eq:gradtrans}
\end{eqnarray}
in eq.~(\ref{eq:octinter}),
it is straightforward to show that~\cite{LS96}
\begin{eqnarray}
{\cal B}_{ijk} & \rightarrow & 
(-)^{\eta_l (\eta_j+\eta_m) +(\eta_l+\eta_m)(\eta_k+\eta_n)} 
\ U_i^{\ l}\  U_j^{\ m}\  U_k^{\ n}\ 
{\cal B}_{lmn}
\ \ \ ,
\label{eq:octtrans}
\end{eqnarray}
where ${\cal B}_{ijk}$ describes a {\bf 70} dimensional representation
of $SU(4|2)_V$. 

It is convenient to decompose the irreducible representations of $SU(4|2)_V$
into irreducible representations of 
$SU(2)_{\rm val}\otimes SU(2)_{\rm   sea}\otimes SU(2)_{\tilde q}$~\cite{CS01a,BBI81,BB81,HM83}.
  The subscript
denotes where the $SU(2)$ acts, either on the $q_V$, the $q_{\rm sea}$, or the
$\tilde q$.  In order to locate a particular baryon in the irreducible
representation we employ the terminology of Ref.~\cite{CS01a}: ground floor,
first floor, second floor and so on, as it is common in the description of
super-algebra multiplets.  The ground floor contains all the baryons that do
not contain a bosonic quark, the first floor contains all baryons that contain
one bosonic quark, the second floor contains all baryons that contain two
bosonic quarks, and the third floor contains the baryons that are comprised
entirely of bosonic quarks.  As a way of distinguishing between baryons
containing some number of valence and sea quarks, we use
``levels''~\cite{CS01a}.  Level A consists of baryons containing no sea quarks,
level B consists of baryons containing one sea quarks, level C consists of
baryons containing two sea quarks, and level D consists of baryons composed
only of sea quarks.

The ground floor of level A of the {\bf 70}-dimensional 
representation contains nucleons, comprised of three 
valence quarks, $q_V q_V q_V$, and is therefore a 
$({\bf 2},{\bf 1},{\bf 1})$ of  
$SU(2)_{\rm val}\otimes SU(2)_{\rm sea}\otimes SU(2)_{\tilde q}$.
The nucleons are embedded as
\begin{eqnarray}
{\cal B}_{abc} & = & {1\over\sqrt{6}}
\left( \ \epsilon_{ab}\ N_c\ +\ 
\epsilon_{ac} N_b\ \right)
\ \ \ ,
\end{eqnarray}
where the indices are restricted to take the values $a,b,c=1,2$ only.
The nucleon doublet is
\begin{eqnarray}
N & = & \left(\matrix{p\cr n}\right)
\ \ \ .
\label{eq:baryons}
\end{eqnarray}
The first floor of level A of the {\bf 70}-dimensional representation 
contains baryons that are composed of two valence quarks and 
one ghost-quark, $\tilde q q_V q_V$, and therefore transforms as 
$({\bf 3}, {\bf 1}, {\bf 2})\oplus ({\bf 1}, {\bf 1}, {\bf 2})$
of $SU(2)_{\rm val}\otimes SU(2)_{\rm sea}\otimes SU(2)_{\tilde q}$ .
The tensor representation $_{\tilde a} \tilde\Se_{ab}$ 
of the $({\bf 3}, {\bf 1}, {\bf 2})$ multiplet, 
and $_{\tilde a} \tilde\Tr$ of the $({\bf 1}, {\bf 1}, {\bf 2})$
multiplet, 
where ${\tilde a}=1,2$ runs over the 
$\tilde q$ indices and $a,b=1,2$ run over the $q_V$ indices,
have baryon assignments
\begin{eqnarray}
_{\tilde a} \tilde\Se_{11} & = & 
\tilde\Sigma_{\tilde a}^{+1}
\ \ ,\ \ 
_{\tilde a} \tilde\Se_{12}\ =\ _{\tilde a} \tilde\Se_{21}\ =\ 
{1\over\sqrt{2}} \tilde\Sigma_{\tilde a}^{0}
\ \ ,\ \ 
_{\tilde a} \tilde\Se_{22}\ =\ 
\tilde\Sigma_{\tilde a}^{-1}
\ \ ,\ \ 
_{\tilde a} \tilde\Tr\ =\ 
 \tilde\Lambda_{\tilde a}^0
\ \ \ .
\label{eq:sixdef}
\end{eqnarray}
The right superscript denotes the third component of $q_V$-isospin,
while the left subscript denotes the $\tilde q$ flavor.
The ground floor of level B of the {\bf 70}-dimensional representation 
contains baryons that are composed of two valence quarks and 
one sea quark, $q_V q_V q_{\rm sea}$, and therefore transforms as 
$({\bf 3}, {\bf 2}, {\bf 1})\oplus ({\bf 1}, {\bf 2}, {\bf 1})$
of $SU(2)_{\rm val}\otimes SU(2)_{\rm sea}\otimes SU(2)_{\tilde q}$ .
The tensor representation $_a \Se_{bc}$ of the 
$({\bf 3}, {\bf 2}, {\bf 1})$ multiplet, 
and $_{a} \Tr$ of the $({\bf 1}, {\bf 2}, {\bf 1})$ multiplet,
where ${a}=1,2$ runs over the 
$q_{\rm sea}$ indices and $b,c=1,2$ run over the $q_V$ indices,
have baryon assignment
\begin{eqnarray}
_{a} \Se_{11} & = & 
\Sigma_{a}^{+1}
\ \ ,\ \ 
_{a} \Se_{12}\ =\ _{a}\Se_{21}\ =\ 
{1\over\sqrt{2}} \Sigma_{a}^{0}
\ \ ,\ \ 
_{a}\Se_{22}\ =\ \Sigma_{a}^{-1}
\ \ ,\ \ 
_{a} \Tr\ =\ \Lambda_{a}^0
\ \ \ .
\label{eq:sixdefsea}
\end{eqnarray}

The $_{\tilde a} \tilde\Se_{ab}$, $_{\tilde a}  \tilde\Tr$, 
$_{a}\Se_{ab}$, and $_{a}\Tr$ are uniquely embedded into
${\cal B}_{ijk}$ (up to field redefinitions), 
constrained by the relations in eq.~(\ref{eq:bianchi}):
\begin{eqnarray}
{\cal B}_{ijk} & = & 
-\sqrt{2\over 3\ }\ _{i-2}\SS_{jk}
\ \ \ \ \ {\rm for}\  \ \ \ i=3,4\ \ {\rm and}\ \ j,k=1,2
\nonumber\\
{\cal B}_{ijk} & = & 
{1\over 2}\ \  _{j-2}\ST\  \varepsilon_{i k }
\ +\ {1\over\sqrt{6}}\ \ _{j-2}\SS_{ik}
\ \ \ \ \ {\rm for}\  \ \ \ j=3,4\ \ {\rm and}\ \ i,k =1,2
\nonumber\\
{\cal B}_{ijk} & = & 
{1\over 2}\ \   _{k-2}\ST\  \varepsilon_{i j }
\ +\ {1\over\sqrt{6}}\ \  _{k-2}\SS_{ij}
\ \ \ \ \ {\rm for}\  \ \ \ k=3,4\ \ {\rm and}\ \ i,j  =1,2
\nonumber\\
{\cal B}_{ijk} & = & 
\sqrt{2\over 3\ }\ _{i-4}\VS_{jk}
\ \ \ \ \ {\rm for}\  \ \ \ i=5,6\ \ {\rm and}\ \ j,k=1,2
\nonumber\\
{\cal B}_{ijk} & = & 
{1\over 2}\ \  _{j-4}\VT\  \varepsilon_{i k }
\ +\ {1\over\sqrt{6}}\ \ _{j-4}\VS_{ik}
\ \ \ \ \ {\rm for}\  \ \ \ j=5,6\ \ {\rm and}\ \ i,k  =1,2
\nonumber\\
{\cal B}_{ijk} & = & 
-{1\over 2}\ \   _{k-4}\VT\  \varepsilon_{ i j }
\ -\ {1\over\sqrt{6}}\ \  _{k-4}\VS_{ij}
\ \ \ \ \ {\rm for}\  \ \ \ k=5,6\ \ {\rm and}\ \ i,j  =1,2
\ \ \ .
\label{eq:embedoctet}
\end{eqnarray}
We do not construct the remaining floors and levels of the {\bf 70}
as we are only interested in one-loop contributions to observables with 
nucleons in the asymptotic states.


As the mass splitting between the 
$\Delta$-resonances and nucleons in QCD is much less than the scale of chiral
symmetry breaking ($\Lambda_\chi\sim 1~{\rm GeV}$)
the $\Delta$-resonances
must be included as a dynamical field in order to have a
theory where the natural scale of higher order interactions is set by
$\Lambda_\chi$.
An interpolating field that contains the spin-${3\over 2}$
$\Delta$-resonances as the ground floor is~\cite{LS96}
\begin{eqnarray}
{\cal T}^{\alpha ,\mu}_{ijk} & \sim &
\left[
Q^{\alpha,a}_i Q^{\beta,b}_j Q^{\gamma,c}_k +
Q^{\beta,b}_i Q^{\gamma,c}_j Q^{\alpha,a}_k  +
Q^{\gamma,c}_i Q^{\alpha,a}_j Q^{\beta,b}_k 
\right]
\varepsilon_{abc} (C\gamma^\mu)_{\beta\gamma}
\ \ \ ,
\label{eq:tdef}
\end{eqnarray}
where the indices $i,j,k$ run from $1$ to $6$.
Neglecting Dirac indices, one finds that
under the interchange of flavor indices~\cite{LS96}
\begin{eqnarray}
{\cal T}_{ijk} & = & 
(-)^{1+\eta_i\eta_j} {\cal T}_{jik}\ =\ 
(-)^{1+\eta_j\eta_k} {\cal T}_{ikj}
\ \ \ .
\label{eq:ttrans}
\end{eqnarray}
${\cal T}_{ijk}$ describes a ${\bf 44}$ dimensional representation
of $SU(4|2)_V$, which has the ground floor of level A transforming as 
$({\bf 4}, {\bf 1}, {\bf 1})$ under 
$SU(2)_{\rm val} \otimes SU(2)_{\rm sea} \otimes SU(2)_{\tilde q}$ with
${\cal T}_{abc}=T_{abc}$,
when the indices are restricted to take the values $a,b,c=1,2$,
and where $T_{abc}$ is the totally symmetric tensor containing
the $\Delta$-resonances,
\begin{eqnarray}
T_{111} & = & \Delta^{++}
\ \ ,\ \ 
T_{112} \ =\  {1\over\sqrt{3}}\Delta^+
\ \ ,\ \ 
T_{122} \ =\  {1\over\sqrt{3}}\Delta^0
\ \ ,\ \ 
T_{222} \ =\   \Delta^{-}
\ \ \ .
\label{eq:decuplet}
\end{eqnarray}
The first floor of level A of the {\bf 44} 
transforms as a $({\bf 3},  {\bf 1}, {\bf 2})$ under
$SU(2)_{\rm val} \otimes SU(2)_{\rm sea} \otimes SU(2)_{\tilde q}$
which has a tensor representation,
$_{\tilde a} \tilde\SD_{ij}$, with baryon assignment
\begin{eqnarray}
_{\tilde a} \tilde\SD_{11} & = & 
\tilde\Sigma_{\tilde a}^{*,+1}
\ \ ,\ \ 
_{\tilde a} \tilde\SD_{12}\ =\ _{\tilde a}\tilde \SD_{21}\ =\ 
{1\over\sqrt{2}} \tilde\Sigma_{\tilde a}^{*,0}
\ \ ,\ \ 
_{\tilde a} \tilde\SD_{22}\ =\ 
\tilde\Sigma_{\tilde a}^{*,-1}
\ \ \ .
\label{eq:sixTdef}
\end{eqnarray}
Similarly,
the ground floor of level B of the {\bf 44} 
transforms as a $({\bf 3}, {\bf 2}, {\bf 1})$ under
$SU(2)_{\rm val} \otimes SU(2)_{\rm sea} \otimes SU(2)_{\tilde q}$
which has a tensor representation,
$_{a} \SD_{ij}$, with baryon assignment
\begin{eqnarray}
_{a} \SD_{11} & = & 
\Sigma_{a}^{*,+1}
\ \ ,\ \ 
_{a} \SD_{12}\ =\ _{a} \SD_{21}\ =\ 
{1\over\sqrt{2}} \Sigma_{a}^{*,0}
\ \ ,\ \ 
_{a} \SD_{22}\ =\ 
\Sigma_{a}^{*,-1}
\ \ \ .
\label{eq:sixTdefsea}
\end{eqnarray}
The embedding of $_{\tilde a} \SD_{ij}$ into ${\cal T}_{ijk}$ is 
unique (up to field redefinitions),
constrained by the symmetry properties in eq.~(\ref{eq:ttrans}):
\begin{eqnarray}
{\cal T}_{ijk} & = & 
+ {1\over\sqrt{3}}\ _{i-2}\SD_{jk}
\ \ \ \ \ {\rm for}\  \ \ \ i=3,4\ \ {\rm and}\ \ j,k=1,2
\nonumber\\
{\cal T}_{ijk} & = & 
+{1\over\sqrt{3}}\ \ _{j-2}\SD_{ik}
\ \ \ \ \ {\rm for}\  \ \ \ j=3,4\ \ {\rm and}\ \ i,k =1,2
\nonumber\\
{\cal T}_{ijk} & = & 
+ {1\over\sqrt{3}}\ \  _{k-2}\SD_{ij}
\ \ \ \ \ {\rm for}\  \ \ \ k=3,4\ \ {\rm and}\ \ i,j =1,2
\nonumber\\
{\cal T}_{ijk} & = & 
+ {1\over\sqrt{3}}\ _{i-4}\tilde \SD_{jk}
\ \ \ \ \ {\rm for}\  \ \ \ i=5,6\ \ {\rm and}\ \ j,k=1,2
\nonumber\\
{\cal T}_{ijk} & = & 
-{1\over\sqrt{3}}\ \ _{j-4}\tilde\SD_{ik}
\ \ \ \ \ {\rm for}\  \ \ \ j=5,6\ \ {\rm and}\ \ i,k =1,2
\nonumber\\
{\cal T}_{ijk} & = & 
+ {1\over\sqrt{3}}\ \  _{k-4}\tilde\SD_{ij}
\ \ \ \ \ {\rm for}\  \ \ \ k=5,6\ \ {\rm and}\ \ i,j =1,2
\ \ \ .
\label{eq:firstfloorT}
\end{eqnarray}
Again, we do not explicitly
construct the second and third floor baryons of the {\bf 44}.

\subsection{Lagrange Density for the Nucleons and $\Delta$-Resonances}

The free Lagrange density for the ${\cal B}_{ijk}$ and 
${\cal T}_{ijk}$ fields is~\cite{LS96}, at LO in the heavy baryon 
expansion~\cite{JMheavy,JMaxial,Jmass,chiralN,chiralUlf},
\begin{eqnarray}
{\cal L} & = & 
i\left(\overline{\cal B} v\cdot {\cal D} {\cal B}\right)
+2\alpha_M^{\rm (PQ)} \left(\overline{\cal B}{\cal B}{\cal M}_+\right)
+2\beta_M^{\rm (PQ)} \left(\overline{\cal B}{\cal M}_+{\cal B}\right)
+2\sigma_M^{\rm (PQ)} \left(\overline{\cal B}{\cal B}\right)\ 
{\rm str}\left({\cal M}_+\right)
\nonumber\\
& - & 
i \left(\overline{\cal T}^\mu v\cdot {\cal D} {\cal T}_\mu\right)
+ 
\Delta\ \left(\overline{\cal T}^\mu {\cal T}_\mu\right)
-2\gamma_M^{\rm (PQ)}
 \left(\overline{\cal T}^\mu{\cal M}_+{\cal T}_\mu\right)
-2 \overline{\sigma}_M^{\rm (PQ)}
  \left(\overline{\cal T}^\mu {\cal T}_\mu\right)\
{\rm str}\left({\cal M}_+\right)
\ ,
\label{eq:free}
\end{eqnarray}
where $\Delta$ is the mass splitting between the ${\bf 70}$ and the ${\bf 44}$,
${\cal M}_+={1\over 2}\left(\xi^\dagger m_Q\xi^\dagger + \xi m_Q\xi\right)$,
and $\xi=\sqrt{\Sigma}$.
The brackets, $\left(\ \right)$ denote contraction of Lorentz and flavor
indices as defined in Ref.~~\cite{LS96}.
For a matrix $\Gamma^\alpha_\beta$ acting in spin-space, 
and a matrix $Y_{ij}$ that acts in flavor-space, 
the required contractions are~\cite{LS96}
\begin{eqnarray}
\left(\overline{\cal B}\  \Gamma \ {\cal B}\right)
& = & 
\overline{\cal B}^{\alpha,kji}\ \Gamma_\alpha^\beta\  {\cal B}_{ijk,\beta}
\ \ ,\ \ 
\left(\overline{\cal T}^\mu\  \Gamma \ {\cal T}_\mu\right)
\ =\ 
\overline{\cal T}^{\mu\alpha,kji}\ \Gamma_\alpha^\beta\  
{\cal T}_{ijk,\beta\mu}
\nonumber\\
\left(\overline{\cal B}\  \Gamma \ Y\ {\cal B}\right)
& = & 
\overline{\cal B}^{\alpha,kji}\ \Gamma_\alpha^\beta\  
Y_i^{\ l}\ 
{\cal B}_{ljk,\beta}
\ \ ,\ \ 
\left(\overline{\cal T}^\mu\  \Gamma \ Y\ {\cal T}_\mu\right)
\ =\ 
\overline{\cal T}^{\mu\alpha,kji}\ \Gamma_\alpha^\beta\  
Y_i^{\ l}\ 
{\cal T}_{ljk,\beta\mu}
\nonumber\\
\left(\overline{\cal B}\  \Gamma \ {\cal B}\ Y\right)
& = & 
(-)^{(\eta_i+\eta_j)(\eta_k+\eta_n)}
\overline{\cal B}^{\alpha,kji}\ \Gamma_\alpha^\beta\  
Y_k^{\ n}\ 
{\cal B}_{ijn,\beta}
\nonumber\\
\left(\overline{\cal B}\  \Gamma \ Y^\mu {\cal T}_\mu\right)
& = & 
\overline{\cal B}^{\alpha,kji}\ \Gamma_\alpha^\beta\  
\left(Y^\mu\right)_i^l
{\cal T}_{ljk,\beta\mu}
\ \ \ ,
\label{eq:Contractions}
\end{eqnarray}
where $\overline{\cal B}$ and $\overline{\cal T}$
transform the same way, e.g.
\begin{eqnarray}
\overline{\cal B}^{kji}&\rightarrow &
(-)^{\eta_l (\eta_j+\eta_m) +(\eta_l+\eta_m)(\eta_k+\eta_n)} 
\ \overline{\cal B}^{nml}\ 
 U_{n}^{\ k\dagger}\ U_{m}^{\ j\dagger}\ U_{l}^{\ i\dagger} \  
\ \ \ .
\end{eqnarray}
The covariant derivative acting on either the ${\cal B}$ or ${\cal T}$ fields
has the form
\begin{eqnarray}
\left({\cal D}^\mu{\cal B}\right)_{ijk} & = & 
\partial^\mu {\cal B}_{ijk}
+
\left(V^\mu\right)^l_i {\cal B}_{ljk}
+ 
(-)^{\eta_i (\eta_j+\eta_m)} \left(V^\mu\right)^m_j {\cal B}_{imk}
+ (-)^{(\eta_i+\eta_j) (\eta_k+\eta_n)}
\left(V^\mu\right)^n_k {\cal B}_{ijn}
\label{eq:covariant}
\end{eqnarray}
where the vector and axial-vector meson fields are
\begin{eqnarray}
V^\mu & = & {1\over 2}\left(\ \xi\partial^\mu\xi^\dagger
\ + \ 
\xi^\dagger\partial^\mu\xi \ \right)
\ \ ,\ \ 
A^\mu \ =\  {i\over 2}\left(\ \xi\partial^\mu\xi^\dagger
\ - \ 
\xi^\dagger\partial^\mu\xi \ \right)
\ \ \ .
\label{eq:mesonfields}
\end{eqnarray}

By restricting ourselves to the $q_V q_V q_V$ sector,
we can make a comparison between the LO 
partially-quenched free Lagrange
density and that of QCD, which has fewer free parameters,
\begin{eqnarray}
{\cal L}^{QCD} & = & 
i\overline{N} v\cdot {\cal D} N
\ +\ 2\alpha_M \overline{N}{\cal M}_+^{\rm QCD} N
\ +\ 2\sigma_M \overline{N} N\ 
{\rm tr}\left[{\cal M}_+^{\rm QCD}\right]
\nonumber\\
& - & 
i \overline{ T}^\mu v\cdot {\cal D} T_\mu
\ +\ 
\Delta\ \overline{T}^\mu T_\mu
\ -\ 2\gamma_M \overline{T}^\mu{\cal M}_+^{\rm QCD} 
 T_\mu
\ -\ 2 \overline{\sigma}_M  \overline{T}^\mu T_\mu\
{\rm tr}\left[{\cal M}_+^{\rm QCD}\right]
\ \ ,
\label{eq:freeQCD}
\end{eqnarray}
where the QCD chirally-invariant mass operator is 
${\cal M}_+^{\rm QCD}=
{1\over 2}\left(\xi^\dagger m_q\xi^\dagger + \xi m_q\xi\right)$,
where $m_q={\rm diag}(m_u,m_d)$, and $\xi$ is the QCD version of 
the matrix defined in eq.~(\ref{eq:phidef}).
The QCD parameters and the PQQCD parameters are related by,
\begin{eqnarray}
\alpha_M & = & {2\over 3}\alpha_M^{(PQ)} - {1\over 3}\beta_M^{(PQ)}
\ \ ,\ \ 
\sigma_M \ =\ \sigma_M^{(PQ)} + {1\over 6}\alpha_M^{(PQ)}
+ {2\over 3}\beta_M^{(PQ)}
\nonumber\\
\gamma_M & = & \gamma_M^{(PQ)}
\ \ ,\ \ 
\overline{\sigma}_M\ =\ \overline{\sigma}_M^{(PQ)}
\ \ \ .
\label{eq:mequals}
\end{eqnarray}

The Lagrange density describing the interactions of the 
${\bf 70}$ and ${\bf 44}$ with the
pseudo-Goldstone bosons at LO in the chiral expansion
is~\cite{LS96},
\begin{eqnarray}
{\cal L} & = & 
2\alpha\ \left(\overline{\cal B} S^\mu {\cal B} A_\mu\right)
\ +\ 
2\beta\ \left(\overline{\cal B} S^\mu A_\mu {\cal B} \right)
\ +\  
2{\cal H} \left(\overline{\cal T}^\nu S^\mu A_\mu {\cal T}_\nu \right)
\nonumber\\
& &  
\ +\ 
\sqrt{3\over 2}{\cal C} 
\left[\ 
\left( \overline{\cal T}^\nu A_\nu {\cal B}\right)\ +\ 
\left(\overline{\cal B} A_\nu {\cal T}^\nu\right)\ \right]
\ ,
\label{eq:ints}
\end{eqnarray}
where $S^\mu$ is the covariant spin-vector~\cite{JMheavy,JMaxial,Jmass}.
Restricting ourselves to the $q_V q_V q_V$ sector, 
we can compare eq.~(\ref{eq:ints}) with the 
LO interaction Lagrange density of QCD,
\begin{eqnarray}
{\cal L} & = & 
2 g_A\  \overline{N} S^\mu  A_\mu N
\ +\ 
g_1\overline{N} S^\mu N\ {\rm tr}
\left[\ A_\mu\ \right]
\ +\  
g_{\Delta N}\ 
\left[\ 
\overline{T}^{abc,\nu}\  A^d_{a,\nu}\  N_b \ \epsilon_{cd} 
\ +\ {\rm h.c.}
\ \right]
\nonumber\\
& & 
+\ 
2 g_{\Delta\Delta}\  
\overline{T}^\nu S^\mu A_\mu T_\nu 
\ +\ 
2 g_{X}\  
\overline{T}^\nu S^\mu \ T_\nu \  {\rm tr}
\left[\ A_\mu\ \right]
\ ,
\label{eq:intsQCD}
\end{eqnarray}
and find that at tree-level
\begin{eqnarray}
\alpha & = & {4\over 3} g_A\ +\ {1\over 3} g_1
\ \ \ ,\ \ \ 
\beta \ =\ {2\over 3} g_1 - {1\over 3} g_A
\ \ \ ,\ \ \ 
{\cal H} \ =\ g_{\Delta\Delta}
\ \ \ ,\ \ \ 
{\cal C} \ =\ -g_{\Delta N}
\ \ \ ,
\label{eq:axrels}
\end{eqnarray}
with $g_X=0$.
Considering only the nucleons, and decomposing the Lagrange density in
eq.~(\ref{eq:intsQCD}) into  the mass eigenstates of the isospin-symmetric
limit, $\pi^{\pm}, \pi^0$ and $\eta$, we have
\begin{eqnarray}
{\cal L} & = & 
2 g_A\  \overline{N} S^\mu  \tilde A_\mu N
\ +\ 
{\sqrt{2}\over f}\ 
\left(g_A+g_1\right) \overline{N} S^\mu N\ \partial_\mu\eta
\ \ ,
\end{eqnarray}
where $\tilde A_\mu$ is the axial-vector field of pions only (excluding the 
isosinglet meson).
In the isospin symmetric limit, with the mass of the $\eta$ being of order 
$\sim\Lambda_{\chi}$, all expressions must be independent of the coupling
$g_1$.

\section{Nucleon Masses}
\label{sec:masses}

\begin{figure}[!ht]
\centerline{{\epsfxsize=4.0in \epsfbox{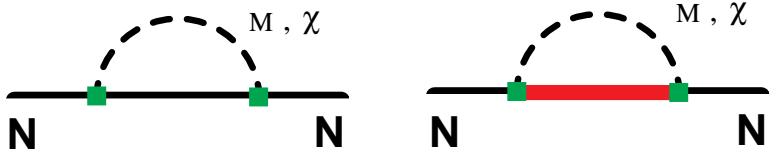}}} 
\vskip 0.15in
\noindent
\caption{\it 
One-loop graphs that give contributions of the form 
$\sim m_Q^{3/2}$
to the masses of the proton and neutron.
A solid, thick-solid and dashed line denote a
{\bf 70}-nucleon, {\bf 44}-resonance, and a meson, respectively.
The solid-squares denote an axial coupling.
}
\label{fig:masses}
\vskip .2in
\end{figure}

The mass of the $i$-th nucleon has a chiral expansion
\begin{eqnarray}
M_i & = & M_0(\mu)\ -\ M_i^{(1)}(\mu)\ -\ M_i^{(3/2)}(\mu)\ +\ ...
\ \ \ ,
\label{eq:massexp}
\end{eqnarray}
where a term $M_i^{(\alpha)}$ denotes a contribution of order $m_Q^\alpha$,
and $i=p,n$. 
The nucleon mass is dominated by a term in the PQ$\chi$PT Lagrange density,
$M_0$, that is independent of $m_Q$.
Each of the contributions 
in eq.~(\ref{eq:massexp})
depends upon the scale chosen to 
renormalize the theory.
While at LO and next-to-leading order (NLO)
in the chiral expansion, the objects $M_0$ and $ M_i^{(1)}$ are scale
independent, at one-loop level they are required to be scale dependent.
The leading dependence upon $m_Q$, occurring at order 
${\cal O}\left(m_Q\right)$, is due to the terms 
in eq.~(\ref{eq:free}) with coefficients $\alpha_M^{(PQ)}$, $\beta_M^{(PQ)}$ 
and $\sigma_M^{(PQ)}$, 
each of which must be determined from lattice simulations.
The leading non-analytic dependence upon $m_Q$ arises from the one-loop
diagrams shown in Fig.~\ref{fig:masses},
and we find the contributions to the proton mass are
\begin{eqnarray}
& & M_p^{(1)} \ =\  
{1\over 3} m_u \left(5\alpha_M^{(PQ)} + 2\beta_M^{(PQ)}\right)
+{1\over 3}  m_d \left(\alpha_M^{(PQ)} + 4\beta_M^{(PQ)}\right)
+2 \sigma_M^{(PQ)}\left(m_j+m_l\right)
\nonumber\\
& & M_p^{(3/2)} \ =\   {1\over 8\pi f^2}\left(\ 
{g_A^2\over 3}\left[\ 
m_{uu}^3+m_{ud}^3+2 m_{ju}^3+ 2 m_{lu}^3 + 3 G_{\eta_u , \eta_u}
\ \right]
\right.\nonumber\\
& & \left. \qquad
\ +\ {g_1^2\over 12}\left[\ 
m_{uu}^3 - 5 m_{ud}^3 + 
3 m_{jd}^3 + 2 m_{ju}^3 + 3 m_{ld}^3 
+ 2 m_{lu}^3 + 3 G_{\eta_u , \eta_u}
+ 6 G_{\eta_u , \eta_d} + 3 G_{\eta_d , \eta_d}
\  \right]
\right.\nonumber\\
& & \left. \qquad
\ +\ {g_A g_1\over 3}\left[\ 
m_{ju}^3 + m_{lu}^3 - m_{ud}^3 + 2 m_{uu}^3 + 3  G_{\eta_u , \eta_d}
+ 3 G_{\eta_u , \eta_u}
\  \right]
\right.\nonumber\\
& & \left. \qquad
+ {g_{\Delta N}^2\over 9\pi}\left[\ 
5 F_{ud} + F_{uu} + F_{ju} + F_{lu} + 2 F_{jd} + 2 F_{ld}
+ 2 E_{\eta_d , \eta_d} +2 E_{\eta_u , \eta_u} 
- 4 E_{\eta_u , \eta_d}
\ \right]\ \right)
\ \ \ ,
\label{eq:pmass}
\end{eqnarray}
where the function $F_{c}=F( m_{c},\Delta,\mu)$ is
\begin{eqnarray}
F (m,\Delta,\mu) & = & 
\left(m^2-\Delta^2\right)\left(
\sqrt{\Delta^2-m^2} \log\left({\Delta -\sqrt{\Delta^2-m^2+i\epsilon}\over
\Delta +\sqrt{\Delta^2-m^2+i\epsilon}}\right)
-\Delta \log\left({m^2\over\mu^2}\right)\ \right)
\nonumber\\
& - & {1\over 2}\Delta m^2 \log\left({m^2\over\mu^2}\right)
\ \ \ .
\label{eq:massfun}
\end{eqnarray}
The functions $G_{\eta_a ,\eta_b}$ and $E_{\eta_a , \eta_b}$ are
$G_{\eta_a ,\eta_b}= 
{\cal H}_{\eta_a\eta_b}(m_{\eta_a}^{3},m_{\eta_b}^{3},m_X^{3})$ 
and 
$E_{\eta_a , \eta_b}={\cal H}_{\eta_a\eta_b}(F_{\eta_a},F_{\eta_b},F_{X})$,
respectively, where the function ${\cal H}_{\eta_a\eta_b}$ is given in 
eq.~(\ref{eq:HPsdef}).
The contributions to the neutron mass are
\begin{eqnarray}
& & M_n^{(1)} \ =\  
{1\over 3} m_u \left(\alpha_M^{(PQ)} + 4\beta_M^{(PQ)}\right)
+{1\over 3}  m_d \left(5\alpha_M^{(PQ)} + 2\beta_M^{(PQ)}\right)
+2 \sigma_M^{(PQ)}\left(m_j+m_l\right)
\nonumber\\
& & M_n^{(3/2)} \ =\  {1\over 8\pi f^2}\left(\ 
{g_A^2\over 3}\left[\ 
m_{dd}^3+m_{ud}^3+2 m_{jd}^3+ 2 m_{ld}^3 + 3 G_{\eta_d , \eta_d}
\ \right]
\right.\nonumber\\
& & \left. \qquad 
+ {g_1^2\over 12}\left[\ 
m_{dd}^3 - 5 m_{ud}^3 + 2 m_{jd}^3 + 3 m_{ju}^3 + 2 m_{ld}^3 + 3 m_{lu}^3
+ 3 G_{\eta_u , \eta_u}
+ 6 G_{\eta_u , \eta_d} + 3 G_{\eta_d , \eta_d}\ \right]
\right.\nonumber\\
& & \left. \qquad 
+ {g_A g_1\over 3}\left[\ 2 m_{dd}^3 - m_{ud}^3 + m_{jd}^3 + m_{ld}^3 
+ 3 G_{\eta_d , \eta_d}+ 3 G_{\eta_u , \eta_d}
\ \right]
\right.\nonumber\\
& & \left. 
\qquad + {g_{\Delta N}^2\over 9\pi}\left[\ 
5 F_{ud} + F_{dd} + F_{jd} + F_{ld} + 2 F_{ju} + 2 F_{lu}
+ 2 E_{\eta_d , \eta_d} +2 E_{\eta_u , \eta_u} 
- 4 E_{\eta_u , \eta_d}
\ \right]\ \right)
\ \ \ .
\label{eq:nmass}
\end{eqnarray}

Our expressions for both the proton and neutron masses collapse down to 
those of isospin-symmetric QCD~\cite{J92} in the limit 
$m_j, m_l, m_u, m_d \rightarrow \overline{m}$,
\begin{eqnarray}
M_N & = & M_0 - 2 \overline{m}\left( \alpha_M+2\sigma_M\right)
- {1\over 8\pi f^2}\left[\ {3\over 2} g_A^2 m_\pi^3
\ +\ {4 g_{\Delta N}^2\over 3\pi} F_\pi\ \right]
\ \ \ .
\end{eqnarray}
In obtaining this result  we have used the fact that 
$G_{\eta_d , \eta_d}\rightarrow -{1\over 2} m_\pi^3$, and 
$E_{\eta_u , \eta_u} \rightarrow -{1\over 2} F_\pi$.
By varying both the sea-quark and valence-quark masses over a suitable range
and determining the proton and neutron masses,
the constants $M_0$, $\alpha_M^{(PQ)}$, $\beta_M^{(PQ)}$ 
and $\sigma_M^{(PQ)}$ can, in principle, be determined for two-flavor QCD.

\section{Nucleon Magnetic Moments}
\label{sec:mm}

The nucleon magnetic moments have been computed on the lattice in
both quenched and unquenched QCD using quark masses
significantly larger than those of nature~\cite{magmoments}.
In order to define the magnetic moments of the nucleons in PQQCD, one must
first define the light-quark electric-charge matrix.
As discussed in detail in Ref.~\cite{CS01a}, the extension of the
electric-charge matrix from QCD, where 
\begin{eqnarray}
{\cal Q} & = & 
{\rm diag}\left(\ +{2\over 3}\ ,\  -{1\over 3}\ \right)
\ \ \ ,
\label{eq:QCDcharge}
\end{eqnarray}
to PQQCD is not unique,
but is constrained by the 
requirement of recovering QCD in the limit that 
$m_j\rightarrow m_u$ and $m_l\rightarrow m_d$.
The most general charge matrix whose matrix elements reduce
down to those of QCD (keeping the valence-quark charges fixed)
is
\begin{eqnarray}
{\cal Q}^{(PQ)} & = & 
{\rm diag}\left(\ +{2\over 3}\ ,\  -{1\over 3}
\ ,\ q_j\ ,\  q_l\ ,\  q_j\ ,\  q_l\ \right)
\ \ \ .
\label{eq:PQcharge}
\end{eqnarray}
For subsequent discussions we define
${\cal Q}_{\xi+}^{(PQ)}  =  {1\over 2} 
\left(\ \xi^\dagger {\cal Q}^{(PQ)}\xi + \xi {\cal Q}^{(PQ)}\xi^\dagger
\right)$, 
and 
${\cal Q}_{\xi+}  =  {1\over 2} 
\left(\ \xi^\dagger {\cal Q}\xi + \xi {\cal Q}\xi^\dagger
\right)$.

Studying the behavior of the magnetic moments for a variety of charge-matrices
with different relative contributions of the singlet and adjoint
representations will yield the desired low-energy constants in the chiral
Lagrangian~\cite{CS01a,GP01a}.  The additional freedom introduced by the
charges $q_l$ and $q_j$ is a blessing rather than a curse due to the
fact that different extensions correspond to different weightings of
disconnected diagrams in lattice simulations.  Thus the impact of disconnected
diagrams whose numerical evaluation, in some cases, converges slowly (and hence
can induce large uncertainties) can be minimized by a
suitable choice of ${\cal Q}^{(PQ)}$.  Furthermore, in PQ$\chi$PT the contribution
of one-loop diagrams involving the more massive sea-quarks can be minimized
(order-by-order) by an appropriate choice of charges, thereby improving the
convergence of the chiral expansion.

In QCD, there are two invariants that can be constructed at LO
in the chiral expansion to describe the magnetic moments of the nucleons, 
\begin{eqnarray}
{\cal L} & = & 
{e\over 4 M_N} F_{\mu\nu}\ 
\left(\ 
\mu_A\ 
{\rm Tr}\left[\ {\cal Q}_{\xi_+}\ \right]
\ \overline{N}\sigma^{\mu\nu} N
\ +\ 
\mu_B\ \overline{N}\sigma^{\mu\nu} 
{\cal Q}_{\xi_+} N
\ \right)
\ \ \ ,
\label{eq:magQCD}
\end{eqnarray}
where $F_{\mu\nu}$ is the electromagnetic field-strength tensor, and
$M_N$ is the physical value of the nucleon mass.
More conventionally, this is  written in terms of the isoscalar and isovector
operators
\begin{eqnarray}
{\cal L} & = & 
{e\over 4 M_N} F_{\mu\nu}\ 
\left(\ 
\mu_0\ \overline{N}\sigma^{\mu\nu} N
\ +\ 
\mu_1\ \overline{N}\sigma^{\mu\nu} 
\tau^3_{\xi+}\ N
\ \right)
\ \ \ ,
\label{eq:magQCDiso}
\end{eqnarray}
where $\mu_0=(2{\mu_A}+{\mu_B})/6$ is the isoscalar nucleon magnetic moment,
$\mu_1={\mu_B}/2$ is the isovector nucleon magnetic moment,
and 
$\tau^a_{\xi+}  =  {1\over 2} 
\left(\ \xi^\dagger\tau^a \xi + \xi\tau^a \xi^\dagger
\right)$.
In PQQCD, the LO Lagrange density contributing to the magnetic moments
of the nucleons has the form 
\begin{eqnarray}
{\cal L} & = & 
{e\over 4 M_N} F_{\mu\nu}\ 
\left[\ 
\mu_\alpha\ \left(\ \overline{\cal B}\ \sigma^{\mu\nu}\  {\cal B}\  
{\cal Q}_{\xi+}^{(PQ)}\ \right)
\ +\ 
\mu_\beta\ \left(\ \overline{\cal B}\ \sigma^{\mu\nu} \ 
{\cal Q}_{\xi+}^{(PQ)}\ 
{\cal B}\ \right)
\right.\nonumber\\
& & \left.\qquad\qquad
+\  \mu_\gamma \ {\rm str}\left[\ {\cal Q}_{\xi+}^{(PQ)}\ \right]
 \left(\ \overline{\cal B}\ \sigma^{\mu\nu}\  {\cal B}\ \right)
\ \right]
\ \ \ .
\label{eq:dimfive}
\end{eqnarray}
It is interesting to note that there is one more operator in PQQCD than there
is in QCD at LO.
The coefficients in the QCD and PQQCD LO Lagrange densities are related by
\begin{eqnarray}
\mu_0 & = & {1\over 6}\ \left[\ \mu_\alpha+\mu_\beta+2\mu_\gamma\ \right]
\ \ ,\ \ 
\mu_1 \ = \ {1\over 6}\ \left[\ 2\mu_\alpha - \mu_\beta\ \right]
\ \ \ .
\label{eq:magrels}
\end{eqnarray}
At the order to which we will be  working, $m_q^{1/2}$,
it is convenient to write the 
proton and neutron magnetic moments in terms of the $\mu_{0,1}$.
However, higher order calculations will need to be performed using
$\mu_{\alpha,\beta,\gamma}$ directly.

Up to order $m_q^{1/2}$, it is convenient to write
the magnetic moment of the $i$-th nucleon as
\begin{eqnarray}
\mu_i & = & \alpha_i\ +\ {M_N\over 4\pi f^2}\ \left[\ 
\beta_i\ +\ \beta_i^\prime\ \right]
\ +\ ...
\ \ \ ,
\label{eq:magdef}
\end{eqnarray}
where the tree-level contributions, $\alpha_i$, are given by the 
Lagrange density in eq.~(\ref{eq:dimfive}).
\begin{figure}[!ht]
\centerline{{\epsfxsize=4.0in \epsfbox{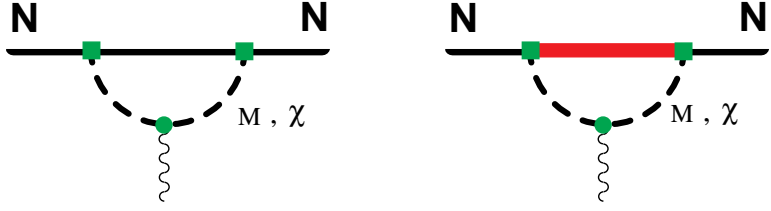}}} 
\vskip 0.15in
\noindent
\caption{\it 
One-loop graphs that give contributions of the form 
$\sim \sqrt{m_Q}$ to the magnetic moments of the nucleons.
A solid, thick-solid and dashed line denote a
{\bf 70}-nucleon, {\bf 44}-resonance, and a meson, respectively.
The solid-squares denote an axial coupling from eq.~(\ref{eq:ints}),
while the solid circle denotes a minimal coupling to the electromagnetic 
field.
}
\label{fig:magmoms}
\vskip .2in
\end{figure}
The $\beta$ and $\beta^\prime$ contributions arise from the one-loop
diagrams shown in Fig.~\ref{fig:magmoms}, involving nucleons in the 
${\bf 70}$ and resonances in the 
${\bf 44}$ dimensional representations, respectively.
For the proton we find
\begin{eqnarray}
\alpha_p & = & \mu_0+\mu_1
\nonumber\\
\beta_p & =  & 
{g_A^2\over 9}\left[\ 4 m_{uu} - 5 m_{ud} - 4 m_{ju} - 4 m_{lu}\ \right]
\ +\ 
{2 g_1 g_A\over 9}\left[\ m_{ud} + m_{uu} - m_{ju} - m_{lu}\ \right]
\nonumber\\ & & 
\ +\ 
{g_1^2\over 36}\left[\ 
m_{ud} + 4 m_{uu} - 3 m_{dd} + 3 m_{jd} - 4 m_{ju} + 3 m_{ld} - 4 m_{lu}\
\right]
\nonumber\\
& & 
+
q_j\left({2 g_A^2\over 3}\left[m_{ju}-m_{uu}\right]
+{g_1 g_A\over 3}\left[m_{ju}-m_{uu}\right]
+{g_1^2\over 6}\left[m_{ju} - m_{uu} + {3\over 2} m_{jd} - 
{3\over 2} m_{ud}\right]
\right)
\nonumber\\
& & 
+ 
q_l\left({2 g_A^2\over 3}\left[m_{lu}-m_{ud}\right]
+{g_1 g_A\over 3}\left[m_{lu}-m_{ud}\right]
+{g_1^2\over 6}\left[m_{lu} - m_{ud} + {3\over 2} m_{ld} - 
{3\over 2} m_{dd}\right]
\right)
\nonumber\\
\beta^\prime_p & = & 
g_{\Delta N}^2\ {1\over 27}\ \left[\ 
\cf_{dd} - \cf_{uu} - 6 \cf_{ud}
- \cf_{jd} + \cf_{ju} - \cf_{ld} + \cf_{lu}
\right.\nonumber\\
& & \left.\qquad
+ {3\over 2}\  q_j\  
\left( \cf_{uu} + 2 \cf_{ud} - \cf_{ju} - 2 \cf_{jd}\right)
+ {3\over 2} \ q_l\  
\left( \cf_{ud} + 2 \cf_{dd} - \cf_{lu} - 2 \cf_{ld}\right)
\ \right]
\ \ \ ,
\label{eq:pmag}
\end{eqnarray}
where the function $\cf_i={\cal F}(m_i,\Delta,\mu)$ is
\begin{eqnarray}
\pi {\cal F}(m,\Delta,\mu)
& = & \sqrt{\Delta^2-m^2}\log\left({\Delta-\sqrt{\Delta^2-m^2+i\epsilon}
\over \Delta+\sqrt{\Delta^2-m^2+i\epsilon}}\right)
\ -\ \Delta\log\left({m^2\over\mu^2}\right)
\ \ \ .
\label{eq:magfun}
\end{eqnarray}
In the limit $\Delta\rightarrow 0$, ${\cal F}(m,0,\mu)=m$.
For the neutron we find
\begin{eqnarray}
\alpha_n & = & \mu_0-\mu_1
\nonumber\\
\beta_n & =  & 
{g_A^2\over 9}\left[7 m_{ud} + 2 m_{ld} + 2 m_{jd} - 2 m_{dd}\right]
+{g_1 g_A\over 9}\left[ m_{jd} + m_{ld} - m_{ud} - m_{dd}\right]
\nonumber\\
& & 
+ {g_1^2\over 18}\left[ 3 m_{uu} + 2 m_{ud} - m_{dd} + m_{jd} - 
3 m_{ju} + m_{ld} - 3 m_{lu}\right]
\nonumber\\
& & 
+ q_j\left( {2 g_A^2\over 3}\left[ m_{jd}-m_{ud}\right]
+{g_1 g_A\over 3}\left[ m_{jd}-m_{ud}\right]
+{g_1^2\over 6} \left[ m_{jd}-m_{ud} +{3\over 2} m_{ju} - 
{3\over 2} m_{uu}\right]\ \right)
\nonumber\\
& & 
+ q_l\left( {2 g_A^2\over 3}\left[ m_{ld}-m_{dd}\right]
+{g_1 g_A\over 3}\left[ m_{ld}-m_{dd}\right]
+{g_1^2\over 6} \left[ m_{ld}-m_{dd} +{3\over 2} m_{lu} - 
{3\over 2} m_{ud}\right]\ \right)
\nonumber\\
\beta^\prime_n & = & 
g_{\Delta N}^2\ {1\over 54}\ \left[\ 
\cf_{dd} - 4 \cf_{uu} + 9 \cf_{ud}
- \cf_{jd} + 4 \cf_{ju} - \cf_{ld} + 4 \cf_{lu}
\right.\nonumber\\
& & \left.\qquad\qquad
+ 3\  q_j\  \left( \cf_{ud} + 2 \cf_{uu} - \cf_{jd} - 2 \cf_{ju}\right)
+ 3 \  q_l\  \left( \cf_{dd} + 2 \cf_{ud} - \cf_{ld} - 2 \cf_{lu}\right)
\ \right]
\ \ .
\label{eq:nmag}
\end{eqnarray}
These expressions reduce to their QCD 
counterparts~\cite{CP74,JLMS92,MS97}
\begin{eqnarray}
\mu_p & = & \mu_0+\mu_1 
- {M_N\over 4\pi f^2}\left[\ g_A^2 \ m_{\pi^+} 
+ {2\over 9}\ g_{\Delta N}^2\  {\cal F}_{\pi^+}\ \right]
\nonumber\\
\mu_n & = & \mu_0-\mu_1 
+ {M_N\over 4\pi f^2}\left[\ g_A^2 \ m_{\pi^+} 
+ {2\over 9}\ g_{\Delta N}^2 \ {\cal F}_{\pi^+}\ \right]
\ \ \ ,
\label{eq:magmomsQCD}
\end{eqnarray}
when $m_j\rightarrow m_u$ and $m_l\rightarrow m_d$. Note that the results are independent of the 
charges $q_{j,l}$.

\section{Matrix Elements of Isovector Twist-2 Operators}
\label{sec:for}

The forward matrix elements of twist-2 operators
play an important role in hadronic structure as they are directly related
to the moments of the parton distribution functions. There exist
both quenched and unquenched lattice simulations of these matrix elements~\cite{gaandtwisttwo}.
Recently, it was realized that the long-distance contributions to 
these matrix elements could be computed
order-by-order in the chiral
expansion using chiral perturbation theory~\cite{AS,CJ,CJb}.
These corrections have been applied to results from both quenched and 
unquenched lattice data~\cite{aussies}, with interesting results.
In addition, the long-distance contributions arising in QQCD
and the large-$N_c$ limit of QCD
have been computed in Ref.~\cite{CSqqcd} and Ref.~\cite{CJNc}, 
respectively.
Further, this technique has been applied to the off-forward
matrix elements of twist-2 operators in order 
to study the spin structure of the  proton~\cite{CJoff}.
In Ref.~\cite{CS01a}, the matrix elements of the isovector twist-2 operators
were computed at the one-loop level in $SU(6|3)_L\otimes SU(6|3)_R$ PQ$\chi$PT.
Like the magnetic moments, the extension of the twist-2 matrix elements
from QCD to PQQCD
introduces a non-unique isovector charge matrix, which can be exploited
to optimize the numerical simulations and the chiral expansion~\cite{CS01a}.

In QCD, the nonsinglet operators have the form,
\begin{eqnarray}
{\cal O}^{ (n), a}_{\mu_1\mu_2\ ... \mu_n}
& = & 
{1\over n!}\ 
\overline{q}\ \tau^a\ \gamma_{ \{\mu_1  } 
\left(i \stackrel{\leftrightarrow}{D}_{\mu_2}\right)\ 
... 
\left(i \stackrel{\leftrightarrow}{D}_{ \mu_n\} }\right)\ q
\ -\ {\rm traces}
\ \ \ ,
\label{eq:twistop}
\end{eqnarray}
where the $\{ ... \}$ denotes symmetrization with respect to all Lorentz indices,
and $\tau^a$ are Pauli matrices acting in flavor-space.
They transform as $({\bf 3},{\bf 1})\oplus  ({\bf 1},{\bf 3})$
under $SU(2)_L\otimes SU(2)_R$ chiral transformations~\cite{AS,CJ}.
Of particular interest to us are the isovector operators where
$\tau^3={\rm diag}(1,-1)$.
At LO in the chiral expansion
the ${\cal O}^{(n), 3}_{\mu_1\mu_2\ ... \mu_n}$
match onto~\cite{AS,CJ,CJb}
\begin{eqnarray}
{\cal O}^{(n),3}_{\mu_1\mu_2 ...\mu_n}
& & \rightarrow 
a^{(n)} \left(i\right)^n {f^2\over 4} 
\left({1\over\Lambda_\chi}\right)^{n-1}
{\rm Tr}\left[\ 
\Sigma^\dagger \tau^3 \overrightarrow\partial_{\mu_1}
 \overrightarrow\partial_{\mu_2}...
 \overrightarrow\partial_{\mu_n}
\Sigma
\ +\ 
\Sigma \tau^3 \overrightarrow\partial_{\mu_1}
 \overrightarrow\partial_{\mu_2}...
 \overrightarrow\partial_{\mu_n}
\Sigma^\dagger \right]
\nonumber\\
& +  & A^{(n)}\ v_{\mu_1} v_{\mu_2}...v_{\mu_n}\ 
\overline{N} \tau^3_{\xi +} N
\nonumber\\
& + &   
\gamma^{(n)} 
\ v_{\mu_1} v_{\mu_2}...v_{\mu_n}\ 
\overline{T}^\alpha\  \tau^3_{\xi +}T_\alpha
\ +\ 
\sigma^{(n)} {1\over n !}
\ v_{\{ \mu_1} v_{\mu_2}...v_{\mu_{n-2}}\ 
\overline{T}_{\mu_{n-1}}\ \tau^3_{\xi +}\ T_{\mu_n\}}
\nonumber\\ & & 
\ -\ {\rm traces}
\ \ \ .
\label{eq:treeQCD}
\end{eqnarray}

In PQQCD, the nonsinglet operators have the form
\begin{eqnarray}
^{PQ}{\cal O}^{(n), a}_{\mu_1\mu_2\ ... \mu_n}
& = & 
{1\over n!}\ 
\overline{Q}\ \overline{\tau}^a\ \gamma_{ \{\mu_1  } 
\left(i \stackrel{\leftrightarrow}{D}_{\mu_2}\right)\ 
... 
\left(i \stackrel{\leftrightarrow}{D}_{ \mu_n\} }\right)\ Q
\ -\ {\rm traces}
\ \ \ ,
\label{eq:Qtwistop}
\end{eqnarray}
where the $\overline{\tau}^a$ are super Pauli matrices,
an extension of the Pauli matrices from two-flavor QCD to PQQCD.
With the requirements that $\overline{\tau}^3$ is supertraceless
and that the QCD matrix elements are recovered
in the limit $m_j\rightarrow m_u$, $m_l\rightarrow m_d$, 
the most general  flavor structure for $\overline{\tau}^3$ is
(keeping the valence-quark charges fixed)
\begin{eqnarray}
\overline{\tau}^3 & = & 
\left(\ 1\  ,\  -1\ ,\  y_j \ ,\  y_l \ ,\   y_j \ ,\  y_l\ \right)
\ \ \ .
\label{eq:isocharge}
\end{eqnarray}
For an arbitrary choice of the $y_i$, this operator contains both 
isovector and isoscalar components.
It is purely isovector only when $y_j+y_l=0$.
As a result, for arbitrary $y_i$, the usual isovector relations between
matrix elements do not hold. 
The fact that disconnected diagrams can only be isoscalar renders this result
obvious.
At LO in the chiral expansion, matrix elements of the isovector
operator
$^{PQ}{\cal O}^{(n), 3}_{\mu_1\mu_2\ ... \mu_n}$
are reproduced by operators of the form~\cite{CS01a,AS,CJ,CJb}
\begin{eqnarray}
^{PQ}{\cal O}^{(n),3}_{\mu_1\mu_2 ...\mu_n}
& & \rightarrow 
a^{(n)} \left(i\right)^n {f^2\over 4} 
\left({1\over\Lambda_\chi}\right)^{n-1}
{\rm str}\left[\ 
\Sigma^\dagger  \overline{\tau}^3 \overrightarrow\partial_{\mu_1}
 \overrightarrow\partial_{\mu_2}...
 \overrightarrow\partial_{\mu_n}
\Sigma
\ +\ 
\Sigma \overline{\tau}^3 \overrightarrow\partial_{\mu_1}
 \overrightarrow\partial_{\mu_2}...
 \overrightarrow\partial_{\mu_n}
\Sigma^\dagger \right]
\nonumber\\
& +  & \alpha^{(n)}\ v_{\mu_1} v_{\mu_2}...v_{\mu_n}\ 
\left(\ \overline{\cal B}\  {\cal B}\  \overline{\tau}^3_{\xi +}\ \right)
\ +\ 
\beta^{(n)}\ v_{\mu_1} v_{\mu_2}...v_{\mu_n}\ 
\left(\ \overline{\cal B}\  \overline{\tau}^3_{\xi +}\  {\cal B}\ \right)
\nonumber\\
& + &   
\gamma^{(n)} 
\ v_{\mu_1} v_{\mu_2}...v_{\mu_n}\ 
\left(\ \overline{\cal T}^\alpha\  \overline{\tau}^3_{\xi +}
\ {\cal T}_\alpha
\right)
\ +\ 
\sigma^{(n)} {1\over n !}
\ v_{\{ \mu_1} v_{\mu_2}...v_{\mu_{n-2}}\ 
\left(\ \overline{\cal T}_{\mu_{n-1}}\  \overline{\tau}^3_{\xi +}\ 
{\cal T}_{\mu_n\}}
\right)
\nonumber\\ & & 
\ -\ {\rm traces}
\ \ \ ,
\label{eq:tree}
\end{eqnarray}
where $ \overline{\tau}^3_{\xi +}\ =\ {1\over 2}\left(
\xi\overline{\tau}^3\xi^\dagger
+\xi^\dagger\overline{\tau}^3\xi\right)$.
In general, the coefficients $a^{(n)}, \ \alpha^{(n)},\ 
\beta^{(n)}, \ \gamma^{(n)}$ and
$\sigma^{(n)}$ are not constrained by symmetries and
must be determined from elsewhere. However for $n=1$ they are
fixed by the isospin charge of the hadrons to be 
\begin{eqnarray}
a^{(1)} & = & +1
\ \ \ ,\ \ \ 
\alpha^{(1)} \ = \ +2
\ \ \ ,\ \ \ 
\beta^{(1)}\ =\ +1
\ \ \ ,\ \ \ 
\gamma^{(1)}\ =\ -3
\ \ \ ,\ \ \ 
\sigma^{(1)}\ =\ 0
\ \ \ .
\end{eqnarray}
Comparing the operators in the QCD and PQQCD chiral Lagrangians,
one finds that
the coefficients $\alpha^{(n)}$ and $\beta^{(n)}$ 
both contribute to $A^{(n)}$ in the QCD limit of PQQCD, 
but away from this limit the operators are independent.

At the one-loop level 
there are contributions from counterterms involving one 
insertion of the quark mass matrix $m_Q$,
\begin{eqnarray}
^{PQ}{\cal O}^{(n),3}_{\mu_1\mu_2 ...\mu_n}
& &\rightarrow 
\left[\ 
b_1^{(n)}\  \cbb^{kji}\ \{\  \overline{\tau}^3_{\xi +}\ ,\ 
{\cal M}_+\ \}^n_i\ \cb_{njk}
\right.\nonumber\\ & & \left.
+\ 
b_2^{(n)}\ (-)^{(\eta_i+\eta_j)(\eta_k+\eta_n)}\ 
\cbb^{kji}\ \{\  \overline{\tau}^3_{\xi +}\ ,\ {\cal M}_+\ \}^n_k\ \cb_{ijn}
\right.\nonumber\\ & & \left.
+\ 
b_3^{(n)}\  (-)^{\eta_l (\eta_j+\eta_n)}\
\cbb^{kji}\  \left(\overline{\tau}^3_{\xi +}\right)^l_i\ 
\left( {\cal M}_+\right)^n_j
\cb_{lnk}
\right.\nonumber\\ & & \left.
+\ 
b_4^{(n)} \  (-)^{\eta_l \eta_j + 1}\ 
\cbb^{kji}\ \left(  
\left(\overline{\tau}^3_{\xi +}\right)^l_i\ \left( {\cal M}_+\right)^n_j
\ +\ \left( {\cal M}_+\right)^l_i 
\left(\overline{\tau}^3\right)^n_j \right)
\cb_{nlk}
\right.\nonumber\\ & & \left.
+\ b_5^{(n)}\  (-)^{\eta_i(\eta_l+\eta_j)}\ 
\cbb^{kji} \left(\overline{\tau}^3_{\xi +}\right)^l_j 
\left( {\cal M}_+\right)^n_i
\cb_{nlk}
+\ b_6^{(n)}\  \cbb^{kji}  \left(\overline{\tau}^3_{\xi +}\right)^l_i 
\cb_{ljk}
\ {\rm str}\left( {\cal M}_+ \right) 
\right.\nonumber\\ & & \left.
+\ b_7^{(n)}\  \ (-)^{(\eta_i+\eta_j)(\eta_k+\eta_n)}\ 
\cbb^{kji}  \left(\overline{\tau}^3_{\xi +}\right)^n_k \cb_{ijn}
\ {\rm str}\left( {\cal M}_+ \right) 
\right.\nonumber\\ & & \left.
+\ b_8^{(n)}\ \cbb^{kji}\ \cb_{ijk} 
\ {\rm str}\left(\overline{\tau}^3_{\xi +}\   {\cal M}_+ \right) 
\ \right]\ v_{\mu_1} v_{\mu_2}...v_{\mu_n}\ 
\ -\ {\rm traces}
\ ,
\label{eq:tcts}
\end{eqnarray}
where the coefficients $b_1^{(n)},...b_8^{(n)}$ are to be determined.
The only constraint that exists on the $b_i^{(n)}$ is that the isospin charge
of each nucleon is absolutely normalized, 
and thus $b_i^{(1)}=0$.

\begin{figure}[!ht]
\centerline{{\epsfxsize=3.0in \epsfbox{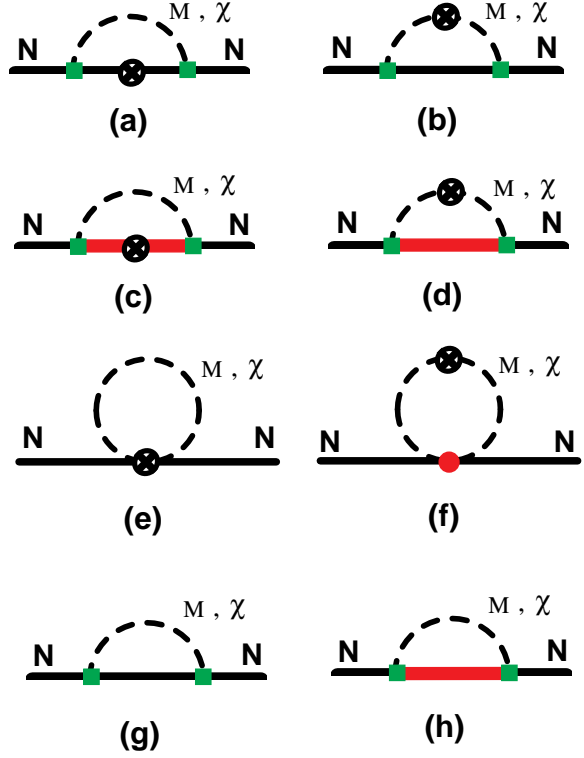}}} 
\vskip 0.15in
\noindent
\caption{\it 
One-loop graphs that give contributions of the form 
$\sim m_Q \log m_Q$
to the matrix elements of the isovector twist-2 operators
in the nucleon.
A solid, thick-solid and dashed line denote a 
{\bf 70}-nucleon, {\bf 44}-resonance, and a meson, respectively.
The solid-squares denote an axial coupling given in eq.(\ref{eq:ints}),
while the solid circle denotes an insertion of the strong two-pion-nucleon
interaction given in eq.(\ref{eq:free}).
The crossed circle denotes an insertion of the tree-level matrix element
of $^{PQ}{\cal O}^{(n), a}_{\mu_1\mu_2\ ... \mu_n}$.
Diagrams (a) to (f) are vertex corrections, while 
diagrams (g) and (h) give rise to wavefunction renormalization.
}
\label{fig:twist}
\vskip .2in
\end{figure}

The forward matrix elements of 
$^{PQ}{\cal O}^{(n),3}_{\mu_1\mu_2 ...\mu_n}$
in the $i$-th nucleon at the one-loop level
can be  written as
\begin{eqnarray}
\langle ^{PQ}{\cal O}^{(n),3}_{\mu_1\mu_2 ...\mu_n} \rangle_i
& = &
\left[\ 
\rho_i^{(n)}
\ +\ {1-\delta^{n1}\over 16\pi^2 f^2}
\left(\ 
\eta_i^{(n), 0}\ -\ \rho_i^{(n)} w_i\ +\ y_{j}\ \eta^{(n), j}_i
\ +\ y_l\ \eta^{(n), l}_i
\ \right)
\right.\nonumber\\ & & \left.\qquad
c_i^{(n), 0} +\ y_{j}\ c^{(n), j}_i
\ +\ y_l\ c^{(n), l}_i
\ \right] \overline{U}_i\  v_{\mu_1} v_{\mu_2}...v_{\mu_n}\ U_i
\ -\ {\rm traces}
\ ,
\label{eq:ttmat}
\end{eqnarray}
where $\rho_i^{(n)}$ are the LO contributions.
The factor of $1-\delta^{n1}$ appears in the higher order corrections
because the isovector charge is not renormalized.
The diagrams shown in Fig.~\ref{fig:twist} give the leading non-analytic
contributions to
the wavefunction renormalization, $w_i$,
the vertex contributions, $\eta_i^{(n), 0}$,
that are independent of
the charges of the ghost- and sea-quarks, and to the vertex contributions,
$\eta_i^{(n), j,l}$, associated with the 
charges of the ghost- and valence-quarks.
For the proton we find
\begin{eqnarray}
\rho_p^{(n)} & = & {1\over 3} \left(\ 2\alpha^{(n)}-\beta^{(n)}\ \right)
\nonumber\\
w_p & = & g_A^2\left( L_{ud}+L_{uu}+2 L_{ju}+2 L_{lu} + 3 R_{\eta_u , \eta_u}
\right)
\nonumber\\
& & 
\ +\ g_1 g_A \left( 2 L_{uu} - L_{ud} + L_{ju} + L_{lu} 
+ 3 R_{\eta_u , \eta_u} + 3 R_{\eta_u , \eta_d}
\right)
\nonumber\\
& & 
\ +\ 
{g_1^2\over 4} \left( L_{uu} - 5 L_{ud} + 2 L_{lu} + 3 L_{ld} + 2 L_{ju} 
+ 3 L_{jd} + 3 R_{\eta_u , \eta_u} + 6 R_{\eta_u , \eta_d}
+ 3 R_{\eta_d , \eta_d}
\right)
\nonumber\\
& & 
+{1\over 3} g_{\Delta N}^2 \left( 5 J_{ud} + J_{uu} + J_{ju}+J_{lu}+2 J_{jd}+2
  J_{ld} + 2 {\cal T}_{\eta_u , \eta_u} + 2 {\cal T}_{\eta_d , \eta_d} 
-4  {\cal T}_{\eta_u , \eta_d} \right)
\nonumber\\
\eta_p^{(n), 0} & = & 3 g_A^2 \rho_p^{(n)} 
\left( L_{uu}-L_{ud} +  R_{\eta_u , \eta_u}\right)
\nonumber\\
& & 
+ {g_1 g_A\over 2}\left[ 
\alpha^{(n)}\left( 3 L_{uu} - L_{ud} + L_{ju} + L_{lu}
+ 4 R_{\eta_u , \eta_d} + 4  R_{\eta_u , \eta_u}\right)
\right.
\nonumber\\
& & \left.
\qquad \qquad 
+ 2 \beta^{(n)} \left(L_{ud} - L_{ju} - L_{lu} - 
R_{\eta_u , \eta_d} - R_{\eta_u , \eta_u}\right)
\right]
\nonumber\\
& & + {g_1^2\over 8}\left[ 
\alpha^{(n)}\left( 3 L_{uu} - 6 L_{ud} - L_{dd} + 5 L_{jd}
+ L_{ju} + 5 L_{ld} + L_{lu} 
\right.\right.\nonumber\\ & & \left. \left.\qquad\qquad
+ 4 R_{\eta_u , \eta_u} + 8  R_{\eta_u , \eta_d}+ 4 R_{\eta_d , \eta_d}
\right)
\right.
\nonumber\\
& & \left.
\qquad
+ 2 \beta^{(n)} \left(
L_{jd} - L_{ju} + L_{ld} - L_{lu} - 2 L_{dd} 
- R_{\eta_u , \eta_u} - 2  R_{\eta_u , \eta_d} -  R_{\eta_d , \eta_d}
\right)
\right]
\nonumber\\
& & 
+ {2\over 9} g_{\Delta N}^2 \left(\gamma^{(n)} - { \sigma^{(n)}\over 3}\right)
\left( J_{dd} - J_{uu} - 6 J_{ud} - 2 J_{jd} - 2 J_{ld} 
- {\cal T}_{\eta_u , \eta_u} - {\cal T}_{\eta_d , \eta_d} 
+ 2 {\cal T}_{\eta_u , \eta_d}
\right)
\nonumber\\
& & 
+{1\over 6} \alpha^{(n)} \left(\ 
5 L_{uu} - 4 L_{ud} - L_{dd} - 5 L_{ju} - 5 L_{lu} + L_{jd} + L_{ld} \right)
\nonumber\\
& & 
+{1\over 3}\beta^{(n)} \left(\ 
L_{uu} + L_{ud} - 2 L_{dd} - L_{ju} - L_{lu} + 2 L_{jd} + 2 L_{ld}\right)
\nonumber\\
\eta_p^{(n), j} & = & g_A^2 \alpha^{(n)} \left( L_{ju}-L_{uu}\right)
\ +\ {g_1 g_A\over 2} \alpha^{(n)}\left( L_{ju}-L_{uu}\right)
\nonumber\\
& &\ +\ {g_1^2\over 8} \left[ 
\alpha^{(n)}\left( L_{jd}+L_{ju} - L_{ud}-L_{uu}\right)
+ \beta^{(n)} \left( 4 L_{jd} + 2 L_{ju} - 4 L_{ud} - 2 L_{uu}\right)
\right]
\nonumber\\
& & + {1\over 9} g_{\Delta N}^2 \left(\gamma^{(n)} 
- { \sigma^{(n)}\over 3}\right)
\left( J_{uu} + 2 J_{ud} - 2 J_{jd} - J_{ju}\right)
\nonumber\\
& & 
+{1\over 6} \alpha^{(n)} \left(\ 
-5 L_{uu} - L_{ud} + 5 L_{ju} + L_{jd}\right)
+{1\over 3}\beta^{(n)} \left(\ 
-L_{uu} - 2 L_{ud} + L_{ju} + 2 L_{jd} \right)
\nonumber\\
\eta_p^{(n), l} & = & g_A^2 \alpha^{(n)} \left( L_{lu}-L_{ud}\right)
\ +\ {g_1 g_A\over 2} \alpha^{(n)}\left( L_{lu}-L_{ud}\right)
\nonumber\\
& &\ +\ {g_1^2\over 8} \left[ 
\alpha^{(n)}\left( L_{ld}+L_{lu} - L_{ud}-L_{dd}\right)
+ \beta^{(n)} \left( 4 L_{ld} + 2 L_{lu} - 4 L_{dd} - 2 L_{ud}\right)
\right]
\nonumber\\
& & 
+ {1\over 9} g_{\Delta N}^2 \left(\gamma^{(n)} - { \sigma^{(n)}\over 3}\right)
\left( J_{ud} + 2 J_{dd} - 2 J_{ld} - J_{lu}\right)
\nonumber\\
& & 
+{1\over 6} \alpha^{(n)} \left(\ 
-5 L_{ud} - L_{dd} + 5 L_{lu} + L_{ld}\right)
+{1\over 3}\beta^{(n)} \left(\ 
-2L_{dd} -  L_{ud} + L_{lu} + 2 L_{ld} \right)
\nonumber\\
c_p^{(n),0} & = & {1\over 3} m_u \left( 2 b_1^{(n)} + 5 b_2^{(n)} - {3\over 2} b_3^{(n)} + 3 b_4^{(n)}
  + 3 b_8^{(n)} \right)\nonumber\\
& &
+ {1\over 3} m_d \left( -4 b_1^{(n)} - b_2^{(n)} + {1\over 2} b_3^{(n)} - 2
  b_4^{(n)} + 2 b_5^{(n)} - 3 b_8^{(n)}
\right)
\nonumber\\
& & 
+ {1\over 3} \left(m_j+m_l\right) \left( -b_6^{(n)}+2 b_7^{(n)} \right)
\nonumber\\
c_p^{(n),j} & = & \left(m_j-m_u\right) b_8^{(n)}
\nonumber\\
c_p^{(n),l} & = & \left(m_l-m_d\right) b_8^{(n)}
\ \ \ ,
\label{eq:ptwist}
\end{eqnarray}
where, for the contributions from the ${\bf 70}$ intermediate states,
we have defined 
$L_{ab} = m_{ab}^2\log\left({m_{ab}^2/\mu^2}\right)$, and 
$R_{x , y} = {\cal H}(L_x, L_y, L_X)$.
For the contributions from the ${\bf 44}$ 
intermediate states, we have defined 
$J_{ab}~=~J(m_{ab}, \Delta, \mu)$, with
\begin{eqnarray}
J(m,\Delta,\mu) & = & 
\left(m^2-2\Delta^2\right)\log\left({m^2\over\mu^2}\right)
+2\Delta\sqrt{\Delta^2-m^2}
\log\left({\Delta-\sqrt{\Delta^2-m^2+ i \epsilon}\over
\Delta+\sqrt{\Delta^2-m^2+ i \epsilon}}\right)
\ \ \ ,
\label{eq:decfun}
\end{eqnarray}
and
${\cal T}_{x , y} = {\cal H}(J_x, J_y, J_X)$.
For the neutron we find 
\begin{eqnarray}
\rho_n^{(n)} & = & -{1\over 3} \left(\ 2\alpha^{(n)}-\beta^{(n)}\ \right)
\nonumber\\
w_n & = & g_A^2\left( L_{dd}+L_{ud}+2 L_{jd}+2 L_{ld} + 3 R_{\eta_d , \eta_d}
\right)
\nonumber\\
& & 
+\ g_1 g_A\left( 2 L_{dd} - L_{ud} + L_{jd} + L_{ld} + 
3 R_{\eta_u , \eta_d} + 3 R_{\eta_d , \eta_d}
\right)
\nonumber\\
& & 
+ {g_1^2\over 4}\left( 
L_{dd} - 5 L_{ud} + 2 L_{jd} + 3 L_{ju} + 2 L_{ld} + 3 L_{lu} 
+ 3 R_{\eta_u , \eta_u} + 6 R_{\eta_u , \eta_d} + 3 R_{\eta_d , \eta_d}
\right)
\nonumber\\
& & 
+{1\over 3} g_{\Delta N}^2 \left( 5 J_{ud} + J_{dd} + 2 J_{ju}+2J_{lu}
+ J_{jd}+  J_{ld} + 2 {\cal T}_{\eta_u , \eta_u} 
+ 2 {\cal T}_{\eta_d , \eta_d} -4  {\cal T}_{\eta_u , \eta_d} \right)
\nonumber\\
\eta_n^{(n), 0} & = & 3 g_A^2 \rho_p^{(n)} 
\left( L_{dd}-L_{ud} +  R_{\eta_d , \eta_d}\right)
\nonumber\\
& & 
+\ {g_1 g_A\over 2}\left[\ \alpha^{(n)}\left(
L_{ud} - 3 L_{dd} - L_{jd} - L_{ld} - 4 R_{\eta_u , \eta_d}
- 4 R_{\eta_d , \eta_d}\right)
\right.
\nonumber\\
& & \left.
\qquad\qquad
+ 2\beta^{(n)}\left( L_{jd} + L_{ld} - L_{ud} 
+R_{\eta_u , \eta_d} + R_{\eta_d , \eta_d} \right)
\right]
\nonumber\\
& & 
+\ {g_1^2\over 8}\left[\alpha^{(n)}\left(6 L_{ud} + L_{uu} - 3 L_{dd} - L_{jd}
    - 5 L_{ju} - L_{ld} - 5 L_{lu} 
\right.\right.\nonumber\\ & & \left. \left.\qquad\qquad
- 4 R_{\eta_u , \eta_u}
- 8  R_{\eta_u , \eta_d} - 4 R_{\eta_d , \eta_d}\right)
\right.
\nonumber\\
& & \left.
\qquad
\ +\ 
2 \beta^{(n)}\left( 
2 L_{uu} +  L_{jd} - L_{ju} + L_{ld} - L_{lu}
+ R_{\eta_u , \eta_u} + 2 R_{\eta_u , \eta_d} +  R_{\eta_d , \eta_d}\right)
\right]
\nonumber\\
& & 
+ {2\over 9} g_{\Delta N}^2 \left(\gamma^{(n)} - { \sigma^{(n)}\over 3}\right)
\left( J_{dd} - J_{uu} + 6 J_{ud} + 2 J_{ju} + 2 J_{lu} 
+ {\cal T}_{\eta_u , \eta_u} + {\cal T}_{\eta_d , \eta_d} 
- 2 {\cal T}_{\eta_u , \eta_d}
\right)
\nonumber\\
& & 
+{1\over 6} \alpha^{(n)} \left(\ 
-5 L_{dd} + 4 L_{ud} + L_{uu} + 5 L_{jd} + 5 L_{ld} - L_{ju} - L_{lu} \right)
\nonumber\\
& & 
+{1\over 3}\beta^{(n)} \left(\ 
-L_{dd} - L_{ud} + 2 L_{uu} - 2 L_{ju} - 2 L_{lu} +  L_{jd} +  L_{ld}\right)
\nonumber\\
\eta_n^{(n), j} & = & g_A^2 \alpha^{(n)} \left( L_{jd}-L_{ud}\right)
\ +\ {g_1 g_A\over 2} \alpha^{(n)} \left( L_{jd}-L_{ud}\right)
\nonumber\\
& & 
+ \ {g_1^2\over 8}\left[\alpha^{(n)}\left(
L_{jd} + L_{ju} - L_{ud} - L_{uu}\right)
+2 \beta^{(n)}\left( L_{jd} + 2 L_{ju} - L_{ud} - 2 L_{uu}\right)\right]
\nonumber\\
& & 
+ {1\over 9} g_{\Delta N}^2 \left(\gamma^{(n)} - { \sigma^{(n)}\over 3}\right)
\left( J_{ud} + 2 J_{uu} - 2 J_{ju} - J_{jd}\right)
\nonumber\\
& & 
+{1\over 6} \alpha^{(n)} \left(\ 
-5 L_{ud} - L_{uu} + 5 L_{jd} + L_{ju}\right)
+{1\over 3}\beta^{(n)} \left(\ 
-L_{ud} - 2 L_{uu} + L_{jd} + 2 L_{ju} \right)
\nonumber\\
\eta_n^{(n), l} & = & g_A^2 \alpha^{(n)} \left( L_{ld}-L_{dd}\right)
\ +\ {g_1 g_A\over 2} \alpha^{(n)} \left( L_{ld}-L_{dd}\right)
\nonumber\\
& & 
+\ {g_1^2\over 8}\left[
\alpha^{(n)}\left(L_{ld} + L_{lu} - L_{dd} - L_{ud}\right)
+2 \beta^{(n)}\left( L_{ld} + 2 L_{lu} - L_{dd} - 2 L_{ud}\right)\right]
\nonumber\\
& & 
+ {1\over 9} g_{\Delta N}^2 \left(\gamma^{(n)} - { \sigma^{(n)}\over 3}\right)
\left( J_{dd} + 2 J_{ud} - 2 J_{lu} - J_{ld}\right)
\nonumber\\
& & 
+{1\over 6} \alpha^{(n)} \left(\ 
-5 L_{dd} - L_{ud} + 5 L_{ld} + L_{lu}\right)
+{1\over 3}\beta^{(n)} \left(\ 
-2L_{ud} -  L_{dd} + L_{ld} + 2 L_{lu} \right)
\nonumber\\
c_n^{(n),0} & = & {1\over 3} m_u 
\left( 4 b_1^{(n)} +  b_2^{(n)} - {1\over 2} b_3^{(n)} + 2 b_4^{(n)} - 2 b_5^{(n)}
  + 3 b_8^{(n)} \right)\nonumber\\
& &
+ {1\over 3} m_d \left( -2 b_1^{(n)} - 5 b_2^{(n)} + {3\over 2} b_3^{(n)} -  3 b_4^{(n)} - 3 b_8^{(n)}
\right)
\nonumber\\
& & 
+ {1\over 3} \left(m_j+m_l\right) \left(b_6^{(n)}-2 b_7^{(n)} \right)
\nonumber\\
c_n^{(n),j} & = & \left(m_j-m_u\right) b_8^{(n)}
\nonumber\\
c_n^{(n),l} & = & \left(m_l-m_d\right) b_8^{(n)}
\ \ \ .
\label{eq:ntwist}
\end{eqnarray}

In the QCD and isospin limits, these expressions reduce down to those obtained
in Refs.~\cite{AS,CJ},
\begin{eqnarray}
\langle {\cal O}^{(n),3}_{\mu_1\mu_2 ...\mu_n} \rangle_p
& = & \overline{U}_p\  v_{\mu_1} v_{\mu_2}...v_{\mu_n} U_p\ 
\left[ \rho_p^{(n)}\left( 1 - {(3 g_A^2+1)(1-\delta^{n1})\over 8\pi^2 f^2}
L_\pi\right)
\right.\nonumber\\ & & \left.\qquad
\ -\ {g_{\Delta N}^2(1-\delta^{n1})\over 4\pi^2 f^2} J_\pi 
\left[ \rho_p^{(n)} + {5\over 9}\gamma^{(n)} - {5\over 27}\sigma^{(n)}\right]
\right]
\ -\ {\rm traces}
\ \ \ ,
\label{eq:ntwistQCD}
\end{eqnarray}
where we have not shown the contribution from local counterterms involving 
a single insertion of $m_q$.
Notice that the QCD limit is independent of the coupling $g_1$, as required.
Furthermore, in the isospin limit alone, where $m_j=m_l$ and 
$m_u=m_d$, these expressions reduce down to those obtained in three-flavor
PQQCD~\cite{CS01a} when the strange quark mass is taken to be very heavy.

\section{The Axial-Vector Current}
\label{sec:axial}

Matrix elements of the axial-vector current, 
$\overline{q}\tau^a\gamma_\mu\gamma_5 q$,
are extensively studied on the lattice~\cite{gaandtwisttwo}.   
In $\chi$PT, there have been numerous computations of these 
matrix elements~\cite{JMaxial,chiralAX} at the one-loop level, 
both including and excluding the decuplet as an intermediate state.
In analogy to the extension 
of the electric- and isovector-charge matrices  
to PQQCD,  there is a non-uniqueness
in the extension of the isovector axial currents,
$\overline{Q}\overline{\tau}^a\gamma_\mu\gamma_5 Q$, to PQQCD.
We use the charge matrix in eq.~(\ref{eq:isocharge}) for the 
flavor-conserving currents, 
and for the flavor-changing current we replace
$\tau^3$ with $\tau^+$ in the upper $2\times 2$ block.
At LO in the chiral expansion, the axial current takes the form

\begin{eqnarray}
^{(PQ)}j_{\mu,5}^3
& &\rightarrow 
2\alpha\ \left(\overline{\cal B} S_\mu {\cal B}\ {\overline{\tau}^3_{\xi +}}\right)
\ +\ 
2\beta\ \left(\overline{\cal B} S_\mu\ {\overline{\tau}^3_{\xi +}}{\cal B} \right)
\ +\  
2{\cal H} \left(\overline{\cal T}^\nu S_\mu\ {\overline{\tau}^3_{\xi +}}{\cal T}_\nu \right)
\nonumber\\
& &  
\ +\ 
\sqrt{3\over 2}{\cal C} 
\left[\ 
\left( \overline{\cal T}_\mu\ {\overline{\tau}^3_{\xi +}} {\cal B}\right)\ +\ 
\left(\overline{\cal B}\ {\overline{\tau}^3_{\xi +}} {\cal T}_\mu\right)\ \right]
\ \ + \ \ldots.
\label{eq:LOaxialcurrent}
\end{eqnarray}

At one-loop level the matrix elements of the axial current 
between nucleons of flavor ``a'' and ``b'' 
are 
\begin{eqnarray}
^{(PQ)}\Gamma_{ab}\ =\ \langle N_b |^{(PQ)}j_{\mu,5} | N_a\rangle
& = & 
\left[\ 
\rho_{ab}
\ +\ {1\over 16\pi^2 f^2}
\left(\ 
\eta_{ab}\ -\ \rho_{ab} {1\over 2}\left[\ w_a+w_b\ \right]
\ +\ y_j\ \eta_{ab}^{(j)}
\ +\ y_l\ \eta_{ab}^{(l)}
\ \right)
\right. \nonumber\\
& & \left. 
\ +\ c_{ab}
\ +\ y_j\ c_{ab}^{(j)}
\ +\ y_l\ c_{ab}^{(l)}
\ \right]
\ 2 \overline{U}_b S_\mu U_a
\ \ \ ,
\label{eq:axmat}
\end{eqnarray}
where the wavefunction renormalization contributions are given in 
eqs.~(\ref{eq:ptwist}) and (\ref{eq:ntwist}).
The $c_{ab}$'s correspond to 
contributions from local counterterms involving one
insertion of the mass matrix, $m_Q$,
\begin{eqnarray}
^{(PQ)}j_{\mu,5}^{m_Q}
& &\rightarrow 
2 \left[\ 
b_1\  \cbb^{kji}\ \{\  \overline{\tau}^3_{\xi +}\ ,\ 
{\cal M}_+\ \}^n_i\ S_\mu \cb_{njk}
\right.\nonumber\\ & & \left.
+\ 
b_2\ (-)^{(\eta_i+\eta_j)(\eta_k+\eta_n)}\ 
\cbb^{kji}\ \{\  \overline{\tau}^3_{\xi +}\ ,\ {\cal M}_+\ \}^n_k\ 
 S_\mu \cb_{ijn}
\right.\nonumber\\ & & \left.
+\ 
b_3\  (-)^{\eta_l (\eta_j+\eta_n)}\
\cbb^{kji}\  \left(\overline{\tau}^3_{\xi +}\right)^l_i\ 
\left( {\cal M}_+\right)^n_j
 S_\mu \cb_{lnk}
\right.\nonumber\\ & & \left.
+\ 
b_4 \  (-)^{\eta_l \eta_j + 1}\ 
\cbb^{kji}\ \left(  
\left(\overline{\tau}^3_{\xi +}\right)^l_i\ \left( {\cal M}_+\right)^n_j
\ +\ \left( {\cal M}_+\right)^l_i 
\left(\overline{\tau}^3\right)^n_j \right)
 S_\mu \cb_{nlk}
\right.\nonumber\\ & & \left.
+\ b_5\  (-)^{\eta_i(\eta_l+\eta_j)}\ 
\cbb^{kji} \left(\overline{\tau}^3_{\xi +}\right)^l_j 
\left( {\cal M}_+\right)^n_i
 S_\mu \cb_{nlk}
\ +\ b_6\  \cbb^{kji}  \left(\overline{\tau}^3_{\xi +}\right)^l_i 
 S_\mu \cb_{ljk}
\ {\rm str}\left( {\cal M}_+ \right) 
\right.\nonumber\\ & & \left.
\ +\ b_7\  \ (-)^{(\eta_i+\eta_j)(\eta_k+\eta_n)}\ 
\cbb^{kji}  \left(\overline{\tau}^3_{\xi +}\right)^n_k 
 S_\mu \cb_{ijn}
\ {\rm str}\left( {\cal M}_+ \right) 
\right.\nonumber\\ & & \left.
+\ b_8\ \cbb^{kji}\  S_\mu \cb_{ijk} 
\ {\rm str}\left(\overline{\tau}^3_{\xi +}\   {\cal M}_+ \right) 
\ \ \right]
\ ,
\label{eq:axcts}
\end{eqnarray}
where the coefficients, $b_i$, must be determined from data or from lattice
calculations.

\begin{figure}[!ht]
\centerline{{\epsfxsize=3.0in \epsfbox{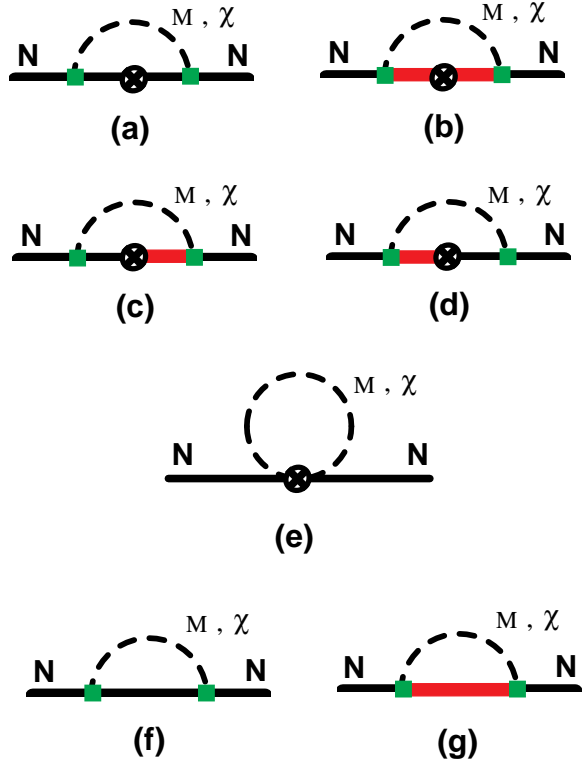}}} 
\vskip 0.15in
\noindent
\caption{\it 
One-loop graphs that give contributions of the form 
$\sim m_Q \log m_Q$ to the matrix elements of 
the axial current in the nucleon.
A solid, thick-solid and dashed line denote a 
{\bf 70}-nucleon, {\bf 44}-resonance, and a meson, respectively.
The solid-squares denote an axial coupling given in eq.(\ref{eq:ints}),
while the crossed circle denotes an insertion of the axial current operator.
Diagrams (a) to (e) are vertex corrections, while 
diagrams (f) and (g) give rise to wavefunction renormalization.
}
\label{fig:ga}
\vskip .2in
\end{figure}

For the matrix element of $\overline{\tau}^3$ between proton states, we find
that the diagrams in Fig.~\ref{fig:ga}
give 
\begin{eqnarray}
\rho_{pp} & = & g_A
\nonumber\\
\eta_{pp} & = & 
g_A^3\ \left[\ L_{ud}-L_{uu}-R_{\eta_u\eta_u}\ \right]
\nonumber\\ & & 
\ + {g_1 g_A^2\over 3}\left[\ L_{ud}-2 L_{uu} - L_{ju}-L_{lu}
- 3 R_{\eta_u\eta_u}- 3 R_{\eta_u\eta_d}\ \right]
\nonumber\\ & & 
\ +{g_A g_1^2\over 12}\left[\ 2 L_{ud} - 4 L_{uu} - 3 L_{jd} 
+ L_{ju} - 3 L_{ld} +     L_{lu} 
- 3 R_{\eta_u\eta_u} - 6 R_{\eta_u\eta_d} -  3 R_{\eta_d\eta_d}\ \right]
\nonumber\\ & & 
\ + {g_1^3\over 24}\left[\ 3 L_{dd}- L_{uu} + 2 L_{ud} - 3 L_{jd} + L_{ju} - 3
  L_{ld} + L_{lu}\ \right]
\nonumber\\ & & 
-\ 
g_A\left[\ L_{ud} - L_{uu} + L_{ju} + L_{lu}\ \right]
\ -\ {g_1\over 2}\left[\ 
L_{dd}-L_{uu} + L_{ju}-L_{jd} + L_{lu}-L_{ld}\ \right]
\nonumber\\ & & 
\ -\ 
{10\over 81} g_{\Delta N}^2 g_{\Delta\Delta}\left[\ 
6 J_{ud}+J_{uu}-J_{dd} + 2 J_{jd}+2 J_{ld} + {\cal T}_{\eta_u\eta_u}
+ {\cal T}_{\eta_d\eta_d}-2{\cal T}_{\eta_u\eta_d}\ \right]
\nonumber\\ & & 
\ +\ 
{8\over 9} g_{\Delta N}^2 g_A \left[\ 
K_{ud} + K_{uu} + K_{ju}+K_{lu} + 2 {\cal S}_{\eta_u\eta_u}
- 2 {\cal S}_{\eta_u\eta_d}\ \right]
\nonumber\\ & & 
\ -\ 
{2\over 9}  g_{\Delta N}^2 g_1\left[\ 
2 K_{dd} - K_{ud} - 3 K_{uu} + 2 K_{jd} - K_{ju} + 2 K_{ld} - K_{lu} 
+ 4 {\cal S}_{\eta_d\eta_d}- 4 {\cal S}_{\eta_u\eta_u}\ \right]
\nonumber\\
\eta_{pp}^{(j)} & = & 
g_A \left( L_{ju}-L_{uu}\right)
\ +\ {g_1\over 2}\left[\ L_{jd}+L_{ju} - L_{ud} - L_{uu}\ \right]
\nonumber\\
& & 
-{5\over 81} g_{\Delta\Delta} g_{\Delta N}^2 \left(
  2 J_{jd} - 2 J_{ud} + J_{ju} - J_{uu}\right)
\nonumber\\
& & 
\ +\ {1\over 3} \left( L_{uu}-L_{ju}\right)
\left[\ {4\over 3} g_A^3 + g_1 g_A^2 + {1\over 4} g_A g_1^2\ \right]
\ +\ {g_1^3\over 72}\left[\ 5 L_{uu} + 9 L_{ud} - 5 L_{ju} - 9 L_{jd}\ \right]
\nonumber\\
& & 
-\ {2\over 9} g_{\Delta N}^2 g_1
\left[\ K_{uu} + 2 K_{ud} - K_{ju} - 2 K_{jd}\ \right]
\nonumber\\
\eta_{pp}^{(l)} & = & 
g_A \left( L_{lu}-L_{ud}\right)
\ +\ {g_1\over 2}\left[\ L_{ld}+L_{lu} - L_{ud} - L_{dd}\ \right]
\nonumber\\
& & 
-{5\over 81} g_{\Delta\Delta} g_{\Delta N}^2 
\left(2 J_{ld} - 2 J_{dd} + J_{lu} - J_{ud}\right)
\nonumber\\
& & 
\ +\ {1\over 3} \left( L_{ud}-L_{lu}\right)
\left[\ {4\over 3} g_A^3 + g_1 g_A^2 + {1\over 4} g_A g_1^2\ \right]
\ +\ {g_1^3\over 72}\left[\ 5 L_{ud} + 9 L_{dd} - 5 L_{lu} - 9 L_{ld}\ \right]
\nonumber\\
& & 
-\ {2\over 9} g_{\Delta N}^2 g_1
\left[\ K_{ud} + 2 K_{dd} - K_{lu} - 2 K_{ld}\ \right]
\nonumber\\
c_{pp} & = & 
m_u\left( {2\over 3} b_1 + {5\over 3} b_2 - {1\over 2} b_3 + b_4 + b_8\right)
\ +\ 
m_d \left( -{4\over 3} b_1 - {1\over 3} b_2 + {1\over 6} b_3 - {2\over 3} b_4
+{2\over 3} b_5 - b_8\right)
\nonumber\\ & & 
\ +\ (m_j+m_l)\left( {2\over 3} b_7 - {1\over 3} b_6\right)
\nonumber\\
c_{pp}^{(j)} & = & b_8 \left( m_j-m_u\right)
\nonumber\\
c_{pp}^{(l)} & = & b_8 \left( m_l-m_d\right)
\ \ \ ,
\label{eq:ppaxial}
\end{eqnarray}
where we have defined the loop function, 
$K_{ab}~=~K(m_{ab},\Delta,\mu)$ with 
\begin{eqnarray}
K(m,\Delta,\mu) & = & 
\left(m^2-{2\over 3}\Delta^2\right)\log\left({m^2\over\mu^2}\right)
\ +\ 
{2\over 3}\Delta \sqrt{\Delta^2-m^2}
\log\left({\Delta-\sqrt{\Delta^2-m^2+ i \epsilon}\over
\Delta+\sqrt{\Delta^2-m^2+ i \epsilon}}\right)
\nonumber\\
& & \ +\ {2\over 3} {m^2\over\Delta} \left(\ \pi m - 
\sqrt{\Delta^2-m^2}
\log\left({\Delta-\sqrt{\Delta^2-m^2+ i \epsilon}\over
\Delta+\sqrt{\Delta^2-m^2+ i \epsilon}}\right)
\right)
\ \ \ ,
\label{eq:Kdecfun}
\end{eqnarray}
and 
${\cal S}_{\eta_a , \eta_b}~=~{\cal H}(K_{aa}, K_{bb}, K_X)$.
For the matrix element of $\overline{\tau}^3$ between neutron states, we find
\begin{eqnarray}
\rho_{nn} & = & -g_A
\nonumber\\
\eta_{nn} & = & 
-g_A^3\ \left[\ L_{ud}-L_{dd}-R_{\eta_d\eta_d}\ \right]
\nonumber\\ & & 
\ + {g_1 g_A^2\over 3}\left[\ 2 L_{dd}- L_{ud} + L_{jd}+L_{ld}
+ 3 R_{\eta_u\eta_d}+ 3 R_{\eta_d\eta_d}\ \right]
\nonumber\\ & & 
\ +{g_A g_1^2\over 12}\left[\ 4 L_{dd} - 2 L_{ud} - L_{jd} 
+ 3 L_{ju} -  L_{ld} + 3 L_{lu} 
+ 3 R_{\eta_u\eta_u} + 6 R_{\eta_u\eta_d} +  3 R_{\eta_d\eta_d}\ \right]
\nonumber\\ & & 
\ + {g_1^3\over 24}\left[\  L_{dd}- 2L_{ud} -3 L_{uu} -  L_{jd} 
+ 3 L_{ju} -   L_{ld} + 3 L_{lu}\ \right]
\nonumber\\ & & 
\ +\ 
g_A\left[\ L_{ud} - L_{dd} + L_{jd} + L_{ld}\ \right]
\ +\ {g_1\over 2}\left[\ L_{uu}-L_{dd} + L_{jd} - L_{ju} 
+ L_{ld}-L_{lu}\ \right]
\nonumber\\ & & 
\ +\ 
{10\over 81} g_{\Delta N}^2 g_{\Delta\Delta}\left[\ 
6 J_{ud}+J_{dd}-J_{uu} + 2 J_{ju}+2 J_{lu} + {\cal T}_{\eta_u\eta_u}
+ {\cal T}_{\eta_d\eta_d}-2{\cal T}_{\eta_u\eta_d}\ \right]
\nonumber\\ & & 
\ -\ 
{8\over 9} g_{\Delta N}^2 g_A \left[\ 
K_{ud} + K_{dd} + K_{jd}+K_{ld} + 2 {\cal S}_{\eta_d\eta_d}
- 2 {\cal S}_{\eta_u\eta_d}\ \right]
\nonumber\\ & & 
\ -\ 
{2\over 9}  g_{\Delta N}^2 g_1\left[\ 
3 K_{dd} + K_{ud} - 2 K_{uu} +  K_{jd} -2  K_{ju} + K_{ld} - 2 K_{lu} 
+ 4 {\cal S}_{\eta_d\eta_d}- 4 {\cal S}_{\eta_u\eta_u}\ \right]
\nonumber\\
\eta_{nn}^{(j)} & = & 
g_A \left( L_{jd}-L_{ud}\right)
\ +\ {g_1\over 2}\left[\ L_{jd}+L_{ju} - L_{ud} - L_{uu}\ \right]
\nonumber\\
& & 
-{5\over 81} g_{\Delta N}^2 g_{\Delta\Delta}
\left( J_{jd}-J_{ud}+2 J_{ju}-2 J_{uu}\right)
\nonumber\\
& & 
\ +\ {1\over 3} \left( L_{ud}-L_{jd}\right)
\left[\ {4\over 3} g_A^3 + g_1 g_A^2 + {1\over 4} g_A g_1^2\ \right]
\ +\ {g_1^3\over 72}\left[\ 5 L_{ud} + 9 L_{uu} - 5 L_{jd} - 9 L_{ju}\ \right]
\nonumber\\
& & 
-\ {2\over 9} g_{\Delta N}^2 g_1
\left[\ K_{ud} + 2 K_{uu} - K_{jd} - 2 K_{ju}\ \right]
\nonumber\\
\eta_{nn}^{(l)} & = & 
g_A \left( L_{ld}-L_{dd}\right)
\ +\ {g_1\over 2}\left[\ L_{ld}+L_{lu} - L_{dd} - L_{ud}\ \right]
\nonumber\\
& & 
-{5\over 81} g_{\Delta N}^2 g_{\Delta\Delta}
\left( J_{ld}-J_{dd}+2 J_{lu}-2 J_{ud}\right)
\nonumber\\
& & 
\ +\ {1\over 3} \left( L_{dd}-L_{ld}\right)
\left[\ {4\over 3} g_A^3 + g_1 g_A^2 + {1\over 4} g_A g_1^2\ \right]
\ +\ {g_1^3\over 72}\left[\ 5 L_{dd} + 9 L_{ud} - 5 L_{ld} - 9 L_{lu}\ \right]
\nonumber\\
& & 
-\ {2\over 9} g_{\Delta N}^2 g_1
\left[\ K_{dd} + 2 K_{ud} - K_{ld} - 2 K_{lu}\ \right]
\nonumber\\
c_{nn} & = & 
-m_u\left( -{4\over 3} b_1 - {1\over 3} b_2 + {1\over 6} b_3 - {2\over 3} b_4
+{2\over 3} b_5 - b_8\right)
\ -\ 
m_d \left( {2\over 3} b_1 + {5\over 3} b_2 - {1\over 2} b_3 + b_4 + b_8\right)
\nonumber\\ & & 
\ -\ (m_j+m_l)\left( {2\over 3} b_7 - {1\over 3} b_6\right)
\nonumber\\
c_{nn}^{(j)} & = & b_8 \left( m_j-m_u\right)
\nonumber\\
c_{nn}^{(l)} & = & b_8 \left( m_l-m_d\right)
\ \ \ .
\label{eq:nnaxial}
\end{eqnarray}
For the $np$ matrix element induced by the 
$\overline{\tau}^+$ axial current, we find 
\begin{eqnarray}
\rho_{np} & = & g_A
\nonumber\\
\eta_{np} & = & 
-g_A^3 R_{\eta_u\eta_d}
\nonumber\\
& & 
+ \left({g_1 g_A^2\over 6}\ +\ {g_1^2 g_A\over 12} \right)
\left[\ 
2 L_{ud} - 2 L_{uu} - 2 L_{dd} - L_{ju}-L_{jd}-L_{lu}-L_{ld}
\right.\nonumber\\ & & \left.\qquad\qquad
- 3  R_{\eta_u\eta_u} - 6  R_{\eta_u\eta_d} - 3 R_{\eta_d\eta_d}
\ \right]
\nonumber\\
& & 
+\ {g_1^3\over 24}
\left[\ 2 L_{ud}+L_{uu}+L_{dd}-L_{ju}-L_{jd}-L_{lu}-L_{ld}\ \right]
\ -\ 
{g_A\over 2}\left[\ L_{ju} + L_{lu} + L_{jd} + L_{ld}\ \right]
\nonumber\\ & & 
\ -\ 
{10\over 81} g_{\Delta N}^2 g_{\Delta\Delta}\left[\ 
4 J_{ud}+J_{uu}+J_{dd} + J_{ju}+ J_{lu} 
+ J_{jd}+ J_{ld} + 2 {\cal T}_{\eta_u\eta_u}
+ 2 {\cal T}_{\eta_d\eta_d}-4{\cal T}_{\eta_u\eta_d}\ \right]
\nonumber\\ & & 
\ +\ 
{4\over 9} g_{\Delta N}^2 g_A \left[\ 
4 K_{ud} + K_{ju}+K_{lu} + K_{jd}+K_{ld} 
+ {\cal S}_{\eta_u\eta_u}
+  {\cal S}_{\eta_d\eta_d}
-2 {\cal S}_{\eta_u\eta_d}\ \right]
\nonumber\\ & & 
\ +\ 
{1\over 9} g_{\Delta N}^2 g_1 \left[\ 
2 K_{ud} +  K_{uu} +  K_{dd} - 
K_{ju}-K_{lu} - K_{jd}-K_{ld} 
\ \right]
\nonumber\\
c_{np} & = & 
(m_u+m_d)
\left( -{1\over 3} b_1 + {2\over 3} b_2 - {1\over 6} b_3 
+{1\over 6} b_4 + {1\over 3} b_5\right)
\ +\ (m_j+m_l)\left( {2\over 3} b_7 - {1\over 3} b_6\right)
\ \ \ ,
\label{eq:npaxial}
\end{eqnarray}
along with 
$\eta_{np}^{(j)} = \eta_{np}^{(l)} = c_{np}^{(j)}= c_{np}^{(l)} = 0$.

These three independent matrix elements each 
reduce down to that of QCD when
$m_j\rightarrow m_u$ and $m_l\rightarrow m_d$.
In the isospin limit and taking $\Delta\rightarrow 0$
the axial-vector current matrix elements in QCD,
at one-loop in the chiral expansion, are~\cite{JMaxial,chiralAX}
\begin{eqnarray}
\Gamma_{np} = g_A\ - \ {L_\pi\over 8\pi^2 f^2}\left(\ 
g_A\left(1+2g_A^2\right) + {2\over 9} g_A g_{\Delta N}^2 + 
{50\over 81} g_{\Delta\Delta} g_{\Delta N}^2
\right)
\ \ \ ,
\end{eqnarray}
where we have not shown the contribution from local counterterms involving one
insertion of $m_q$.
We have used the fact that $K_x, J_x\rightarrow L_x$ 
in the $\Delta\rightarrow 0$ limit.

\section{Conclusions}
\label{sec:conc}

The properties of the neutron and the proton have provided an important benchmark
for comparing the predictions of continuum hadronic effective field theories
(and models) with
nature.  They now provide a benchmark for the ever increasing number of lattice
QCD calculations, be they quenched, partially-quenched or unquenched.
In this paper we have analyzed several observables in two-flavor PQ$\chi$PT to allow for
extrapolations of current and future lattice simulations of two-flavor QCD.
One-loop level computations of the nucleon masses, magnetic moments,
axial-vector currents, as well as the forward matrix elements of isovector
twist-2 operators that are directly related to parton distribution functions,
have been presented.

\bigskip\bigskip

\acknowledgements

We thank Paul Rakow for useful discussions.
This work is supported in part by the U.S. Dept. of Energy under Grants No. DE-FG03-97ER4014.


\begin{references}

\bibitem{Sharpe90}
S.R.~Sharpe,
{\it Nucl. Phys.}  {\bf B17} (Proc. Suppl.), 146 (1990).

\bibitem{S92}
S.R.~Sharpe, 
{\it Phys. Rev.} {\bf D46}, 3146 (1992). 

\bibitem{BG92}
C.~Bernard and M.F.L.~Golterman,
{\it Phys. Rev.}  {\bf D46}, 853 (1992).

\bibitem{LS96}
J.N.~Labrenz and S.R.~Sharpe,
{\it Phys. Rev.}  {\bf D54}, 4595 (1996).

\bibitem{S01a}
M.J.~Savage, {\it Nucl. Phys.} {\bf A700}, 359 (2002).

\bibitem{Pqqcd}
S.R.~Sharpe and N.~Shoresh,
{\it Phys. Rev.} {\bf D62}, 094503 (2000);
{\it Nucl. Phys. Proc. Suppl.} {\bf 83}, 968 (2000);
M.F.L.~Golterman and K.-C.~Leung,
{\it Phys. Rev.} {\bf D57}, 5703 (1998).
S.R.~Sharpe,
{\it Phys. Rev.} {\bf D56}, 7052 (1997);
C.W.~Bernard and M.F.L.~Golterman,
{\it Phys. Rev.} {\bf D49}, 486 (1994).

\bibitem{SS01}
S.R.~Sharpe and N.~Shoresh, {\it Int. J. Mod. Phys.} {\bf A16s1c}, 1219 (2001);
{\it Phys. Rev.} {\bf D64}, 114510 (2001).

\bibitem{CS01a}
J.-W.~Chen and M.J.~Savage,
{\tt hep-lat/0111050}.

\bibitem{BS02a}
S.R.~Beane and M.J.~ Savage,
{\tt hep-lat/0202013}.


\bibitem{BBI81}
A.B.~Balantekin, I.~Bars and F.~Iachello,
{\it Phys. Rev. Lett.} {\bf 47}, 19 (1981).

\bibitem{BB81}
 A.B.~Balantekin and I.~Bars,
{\it J. Math. Phys.} {\bf 23}, 1239 (1981);
{\it J. Math. Phys.} {\bf 22}, 1810 (1981);
{\it J. Math. Phys.} {\bf 22}, 1149 (1981).

\bibitem{HM83}
J.-P.~Hurni and B.~Morel,
{\it J. Math. Phys.} {\bf 24}, 157 (1983).

\bibitem{JMheavy}
E.~Jenkins and  A.V.~Manohar,
{\it Phys. Lett.} {\bf B255}, 558 (1991). 

\bibitem{JMaxial}
E.~Jenkins and  A.V.~ Manohar,
{\it Phys. Lett.} {\bf B259}, 353 (1991). 

\bibitem{Jmass}
E.~Jenkins,
{\it Nucl. Phys.} {\bf B368}, 190 (1992). 

\bibitem{chiralN}
E.~Jenkins and A.V.~Manohar,
{\it Baryon Chiral Perturbation Theory},
talks presented at the workshop on {\it Effective Field Theories
of the Standard Model}, Dobogoko, Hungary (1991);

\bibitem{chiralUlf}
For a recent review see
U.-G.~Mei\ss ner,
Essay for the Festschrift in honor of Boris Ioffe, 
in ``Encyclopedia of Analytic QCD'', 
edited by M. Shifman, World Scientific, ISBN-981-02-4968-3. 

\bibitem{J92}
E.~Jenkins,
{\it Nucl. Phys.} {\bf B368}, 190 (1992);
V.~Bernard, N.~Kaiser and U.-G.~Mei\ss ner,
{\it Z. Phys.} {\bf C60}, 111 (1993).


\bibitem{magmoments}
T.~Draper, R.M.~Woloshyn and K.F.~Liu, 
{\it Phys. Lett.} {\bf 234}, 121 (1990);
D.B.~Leinweber, R.M.~Woloshyn and T.~Draper, 
{\it Phys. Rev.} {\bf D43}, 1659 (1991);
W.~Wilcox, T.~Draper and K.F.~Liu, 
{\it Phys. Rev.} {\bf D46}, 1109 (1992); 
S.~Capitani {\it et al}, {\tt hep-lat/9711007}; 
S.J.~Dong, K.F.~Liu and A.G.~Williams, 
{\it Phys. Rev.}  {\bf D58}, 074504 (1998); 
A.G.~Williams, {\it Nucl. Phys. Proc. Suppl.} {\bf 73}, 306 (1999);
V.~Gadiyak, X.D.~Ji and C.W.~Jung,
{\it Nucl. Phys. Proc. Suppl.}  {\bf 106}, 296 (2002).


\bibitem{GP01a}
M.F.L.~Golterman and E.~Pallante,
{\it JHEP} {\bf  0110}, 037 (2001); 
{\it Nucl. Phys. Proc. Suppl.} {\bf  106}, 335 (2002); 
{\tt hep-lat/0108029}.

\bibitem{CP74}D. G.~Caldi and H.~Pagels,
{\it Phys. Rev.} {\bf D10}, 3739 (1974).

\bibitem{JLMS92}
E.~Jenkins, M.~Luke, A.V.~Manohar and M.J.~Savage,
{\it Phys. Lett.} {\bf B302}, 482 (1993); {\bf B388}, 866 (1996)(E).


\bibitem{MS97}
U.-G.~Mei\ss ner and S.~Steininger  
{\it Nucl. Phys.}  {\bf B499}, 349 (1997).


\bibitem{gaandtwisttwo}
M.~Fukugita, Y.~Kuramashi, M.~Okawa and A.~Ukawa,
{\it Phys. Rev. Lett.}  {\bf 75}, 2092 (1995);
M.~Gockeler {\it et al},
{\it Phys. Rev.}  {\bf D53}, 2317 (1996);
M.~Gockeler {\it et al}
{\it Nucl. Phys. Proc. Suppl.}  {\bf 53}, 81 (1997);
C.~Best {\it et al.}, {\tt hep-ph/9706502};
R.~Horsley  [UKQCD Collaboration],
{\it Nucl. Phys. Proc. Suppl.}  {\bf 94}, 307 (2001);
T.~Blum, S.~Ohta and S.~Sasaki,
{\it Nucl. Phys. Proc. Suppl.}  {\bf 94}, 295 (2001);
D.~Dolgov {\it et al.},
{\it Nucl. Phys. Proc. Suppl.}  {\bf 94}, 303 (2001)
S.~Sasaki, T.~Blum, S.~Ohta and K.~Orginos  [RBC Collaboration],
{\it Nucl. Phys. Proc. Suppl.}  {\bf 106}, 302 (2002);
M.~Gockeler, R.~Horsley, D.~Pleiter, P.E.~Rakow and G.~Schierholz,
{\tt hep-ph/0108105};
D.~Dolgov {\it et al.}  [LHPC collaboration],
{\tt hep-lat/0201021}.


\bibitem{AS}
D.~Arndt and M.J.~Savage, 
{\it Nucl.Phys.} {\bf A697}, 429 (2002). 

\bibitem{CJ}
J-.W.~Chen and X.~Ji,
{\tt hep-ph/0105197}.

\bibitem{CJb}
J-.W.~Chen and X.~Ji,
{\it Phys. Rev. Lett.} {\bf 87}, 152002 (2001).

\bibitem{aussies}
W.~Detmold, W.~Melnitchouk, J.W.~Negele, D.B.~Renner, and  A.W.~Thomas,
{\it Phys. Rev. Lett.} {\bf 87}, 172001 (2001). 

\bibitem{CSqqcd}
J.-W.~Chen and M.J.~Savage,
{\tt nucl-th/0108042}.

\bibitem{CJNc}
J-.W.~Chen and X.~Ji,
{\tt hep-ph/0105296}.

\bibitem{CJoff}
J-.W.~Chen and X.~Ji,
{\it Phys. Rev. Lett.} {\bf 88}, 052003 (2002). 

\bibitem{chiralAX}
J.~Bijnens, H.~Sonoda and M.B.~Wise,
{\it Nucl. Phys.} {\bf B261}, 185 (1985);
M.N.~Butler, R.P.~Springer and M.J.~Savage,
{\it Nucl. Phys.} {\bf B399}, 69 (1993);
M.A.~Luty and M.J.~White,
{\it Phys. Lett.} {\bf B319}, 261 (1993);
{\tt hep-ph/9304291};
R.P.~Springer, {\tt hep-ph/9508324};
M.~Kim and S.~Kim,
{\it Phys. Rev. } {\bf D58}, 074509 (1998);
M.J.~Savage and J.~Walden,
{\it Phys. Rev.} {\bf D55}, 5376 (1997);
R.~Flores-Mendieta, E.~Jenkins and A.V.~Manohar,
{\it Phys. Rev. } {\bf D58}, 094028 (1998);
B.~Borasoy,
{\it Phys. Rev.} {\bf D59}, 054021 (1999);
B.R.~Holstein,
{\it Few. Body. Syst. Suppl.} {\bf 11},116 (1999);
S.-L.~Zhu, S.~Puglia and M.J.~Ramsey-Musolf,
{\it Phys. Rev.} {\bf D63}, 034002 (2001);


\end{references}
\end{document}